\documentclass[lettersize,journal]{IEEEtran}
\usepackage{amsmath}
\usepackage{amsfonts}
\usepackage{amssymb}
\usepackage{mathtools}
\usepackage{algorithmic}
\usepackage{algorithm}
\usepackage{array}
\usepackage{pifont}
\usepackage[table]{xcolor}
\usepackage{tabularx}
\usepackage{makecell}
\usepackage{multirow}
\usepackage[compact]{titlesec}
\usepackage[caption=false,font=normalsize,labelfont=sf,textfont=sf]{subfig}
\usepackage{textcomp}
\usepackage{stfloats}
\usepackage{url}
\usepackage{verbatim}
\usepackage{graphicx}
\usepackage{cite}
\usepackage{dsfont}
\usepackage[hidelinks]{hyperref}
\hypersetup{
  colorlinks = true, 
  citecolor = black, 
  linkcolor = black, 
  urlcolor = black
}

\newcommand{\RNum}[1]{\uppercase\expandafter{\romannumeral #1\relax}}
\DeclareMathOperator*{\argmax}{argmax}

\newcommand{\xmark}{\ding{53}}
\definecolor{gray}{HTML}{E0E0E0}

\newcommand{\rw}[1]{\textcolor{black}{#1}}

\begin{document}

\title{Adaptive Bi-Recommendation and Self-Improving Network for Heterogeneous Domain Adaptation-Assisted IoT Intrusion Detection}

\author{Jiashu Wu,~\IEEEmembership{Graduate Student Member,~IEEE},
        Yang Wang,~\IEEEmembership{Member,~IEEE}\thanks{* Yang Wang is the corresponding author},
        Hao Dai,~\IEEEmembership{Graduate Student Member,~IEEE},
        Chengzhong Xu,~\IEEEmembership{Fellow,~IEEE},
        and Kenneth B. Kent,~\IEEEmembership{Senior Member,~IEEE}
\thanks{Jiashu Wu, Yang Wang and Hao Dai are with Shenzhen Institute of Advanced Technology, Chinese Academy of Sciences, Shenzhen 518055, China. Email: \{js.wu, yang.wang1, hao.dai\}@siat.ac.cn}
\thanks{Jiashu Wu and Hao Dai are also with University of Chinese Academy of Sciences, Beijing 100049, China.}
\thanks{Chengzhong Xu is with the State Key Laboratory of IoT for Smart City, Faculty of Science and Technology, University of Macau, Macau 999078, China. Email: czxu@um.edu.mo}
\thanks{Kenneth B. Kent is with University of New Brunswick, Fredericton, New Brunswick E3B 5A3, Canada. Email: ken@unb.ca}
\thanks{Manuscript received January 1, 2022; revised January 1, 2023.}
\thanks{Copyright (c) 2023 IEEE. Personal use of this material is permitted. However, permission to use this material for any other purposes must be obtained from the IEEE by sending a request to pubs-permissions@ieee.org. }}

\markboth{IEEE Internet of Things Journal}%
{Wu \MakeLowercase{\textit{et al.}}: Adaptive Bi-Recommendation and Self-Improving Network for Heterogeneous Domain Adaptation-Assisted IoT Intrusion Detection}


\maketitle

\begin{abstract}
As Internet of Things devices become prevalent, using intrusion detection to protect IoT from malicious intrusions is of vital importance. However, the data scarcity of IoT hinders the effectiveness of traditional intrusion detection methods. To tackle this issue, in this paper, we propose the Adaptive Bi-Recommendation and Self-Improving Network (ABRSI) based on unsupervised heterogeneous domain adaptation (HDA). The ABRSI \rw{transfers enrich} intrusion knowledge \rw{from a} data-rich network intrusion source domain to facilitate effective intrusion detection for data-scarce IoT target domains. The ABRSI achieves fine-grained intrusion knowledge transfer via adaptive bi-recommendation matching. Matching the bi-recommendation interests of two recommender systems and the alignment of intrusion categories in the shared feature space form a mutual-benefit loop. Besides, the ABRSI uses a self-improving mechanism, autonomously improving the intrusion knowledge transfer from four ways. A hard pseudo label voting mechanism jointly considers recommender system decision and label relationship information to promote more accurate hard pseudo label assignment. To promote diversity and target data participation during intrusion knowledge transfer, target instances \rw{failing to be} assigned with \rw{a hard} pseudo label will be assigned with a probabilistic soft pseudo label, forming a hybrid pseudo-labelling strategy. Meanwhile, the ABRSI also makes soft pseudo-\rw{labels} globally diverse and individually certain. Finally, an error knowledge learning mechanism is utilised to adversarially exploit factors that causes detection ambiguity and learns through both current and previous error knowledge, preventing error knowledge forgetfulness. Holistically, these mechanisms form the ABRSI model that boosts IoT intrusion detection accuracy via HDA-assisted intrusion knowledge transfer. Comprehensive experiments on several intrusion datasets demonstrate the state-of-the-art performance of the ABRSI method, outperforming its counterparts by $9.2\%$, and also verify the effectiveness of ABRSI constituting components and ABRSI's overall efficiency. 
\end{abstract}


\section{Introduction}\label{sec:sec_introduction}

\begin{figure}[!t]
  \begin{center}
    \includegraphics[width=0.48\textwidth,keepaspectratio]{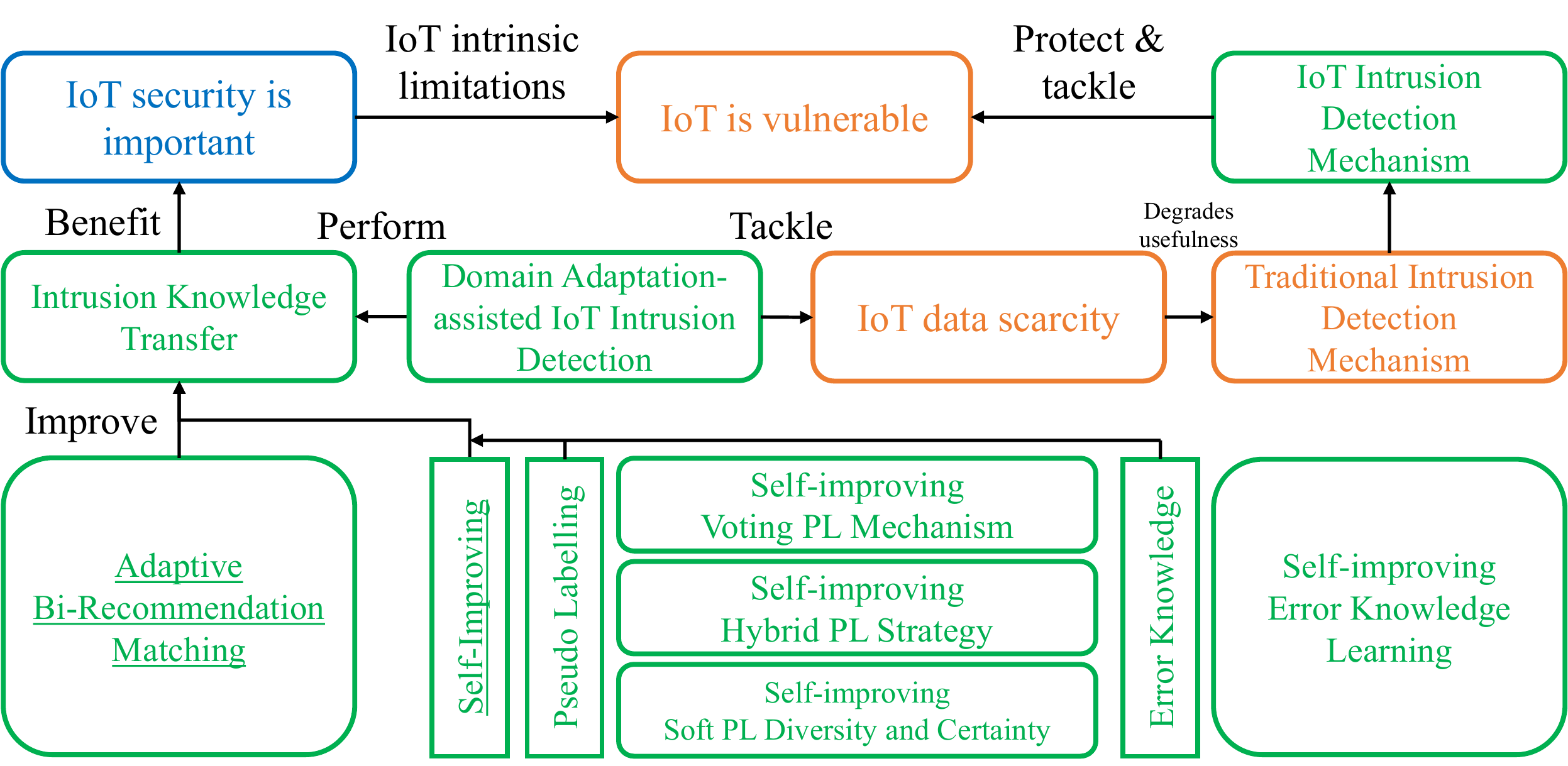}\\
    \caption{The intuition of the ABRSI model}
    \vspace{-0.5cm}
    \label{fig:figure_idea}
  \end{center}
\end{figure}

\IEEEPARstart{A}{s} Internet of Things (IoT) devices evolve rapidly, they transform various applications to become more intelligent \cite{zhang2022numerical,10049541,zhao2023vector}. However, IoT tends to be computational and energy-constrained, which prevents some powerful intrusion detection mechanisms from being deployed, together with a lack of regular maintenance, these limitations make \rw{IoT devices} vulnerable to intrusions \cite{lu2018internet}. Hence, an effective intrusion detection mechanism is necessary to protect IoT infrastructures from malicious intrusions. 

Previous research efforts attempted to tackle the IoT intrusion detection (IID) problem mainly via the rule-based and machine learning-based perspectives. For instance, Chen et al., \cite{jun2014design} and Dietz et al., \cite{dietz2018iot} proposed to use complex event processing and a proactive violation scanning mechanism to detect events that match violation patterns defined in a rule repository. On the other hand, intrusion detectors based on ML models such as stacked autoencoder \cite{muhammad2020stacked}, multi-kernel SVM \cite{murali2019lightweight} and capsule network \cite{yao2019capsule} etc., have been proposed and achieved satisfying performance. However, these efforts either require a sophisticated rule repository which is highly expertise-dependent and can hardly be thorough, or depend on a fully-annotated training dataset, which is labour-intensive to build. Moreover, the limited storage and communication capability and the involvement of user privacy make IoT data hard to be obtained and publicly released. Together with the cost of expertise and manual annotation, IoT tends to be data scarce \cite{9933783}, hindering the usefulness of rule and ML-based methods under IoT scenarios. 

Recently, domain adaptation-based (DA) intrusion detection methods are proposed, which transfer intrusion knowledge \rw{from the} data-rich network intrusion (NI) domain to facilitate more effective IoT intrusion (II) detection under data scarcity. By aligning both domains into a common feature space, the enriched intrusion knowledge from the source NI domain can benefit the target II intrusion detection. Besides, a heterogeneous domain adaptation-based (HDA) method \cite{10026337} should also be able to tackle inter-domain heterogeneities, such as different feature representations, different feature dimensions, etc. For example, Hu et al., \cite{hu2022deep} proposed a deep subdomain adaptation network which transferred intrusion knowledge via local maximum mean discrepancy minimisation. Xie et al., \cite{xie2022collaborative} presented a collaborative alignment framework, which reduced global domain discrepancy and preserves local semantic consistency. Lv et al., \cite{liang2021pareto} put forward the Pareto domain adaptation method that controls the overall optimisation direction for better intrusion knowledge transfer. 

Despite the effectiveness of these HDA-based methods, they still have some limitations that need to be addressed. 
\begin{itemize}
  \item Firstly, recommender system (RS) is an effective way to exploit ``interest'', i.e., intrusion knowledge of intrusion categories. However, to our best knowledge, RS and bi-recommendation matching haven't been introduced to improve HDA-based intrusion detection. 
  \item Secondly, pseudo label (PL) can exploit the potential of unlabeled target II domain. However, to our best knowledge, none of the previous method assigned PL based on the knowledge gained from RS due to their complete ignorance of the RS. Besides, previous methods assigned PL in an isolated manner without considering label relationship information. 
  \item Thirdly, when utilising PL, previous methods either use it \rw{in a} one-hot way, i.e., hard PL \cite{long2018conditional}, or assign a probabilistic soft PL \cite{liang2021pareto,jin2020minimum}. These two ways either hurt participation and diversity of target II instances during knowledge transfer, or lacked enough emphasis of confident and correct PL. None of them considered a hybrid PL strategy that jointly involves both hard and soft PL for better participation, diversity and enough emphasis of confident and correct predictions. 
  \item Finally, although some methods performed intrusion knowledge transfer via minimising the divergence of intrusion category probabilistic correlation semantics \cite{li2019joint}, i.e., error knowledge, they didn't learn it from an adversarial manner and simultaneously consider both the current and past error knowledge, resulting in error knowledge forgetfulness. 
\end{itemize}

\rw{To address these limitations and enhance the intrusion knowledge transfer, by following the high level intuition illustrated in Fig. \ref{fig:figure_idea}, }we propose an Adaptive Bi-Recommendation and Self-Improving (ABRSI) \rw{network in an} unsupervised HDA setting, i.e., the target II domain is completely unsupervised, meanwhile heterogeneities \rw{are present} between source NI and target II domains. To mine the intrusion category interest and benefit intrusion knowledge transfer, we utilise an Adaptive Bi-Recommendation (ABR) matching mechanism. The ABR leverages two RSs, one for each domain, and matches the recommendation for each intrusion category between two RSs, which enforces feature alignment in the shared feature space. Better feature alignment in turn can produce enhanced RS. By adaptively iterating this learning process, it forms a mutually-beneficial positive loop. Then, the ABRSI method also involves Self-Improving (SI), which autonomously improves the intrusion knowledge \rw{transfer in} four ways. Firstly, we jointly consider the neural network prediction, recommender system decision and label relationship information to improve PL accuracy and prevent negative transfer caused by error-prone PLs. A hard PL will only be assigned if these votes reach a consensus, promoting confident and correct PL assignment as much as possible. Secondly, the hard-PL-only strategy prevents unlabeled target instances from participating the intrusion knowledge transfer if a hard PL is not assignable. The diversity of hard PL may also be impaired as hard PLs tend to focus on easy intrusion categories. Thus, we utilise a hybrid PL strategy that involves both hard and probabilistic soft PL to emphasise confident and correct predictions, and meanwhile improve participation and diversity. Thirdly, we use the Tsallis entropy and a diversity maximisation mechanism, aiming for globally diverse and individually certain soft PL. Finally, the ABRSI applies an error knowledge learning (EKL) mechanism, which emphasises error knowledge, i.e., factors that cause prediction ambiguities, and tackling it in an adversarial manner. Meanwhile, both current and previous error knowledge are involved to prevent error knowledge forgetfulness. Holistically, it forms the ABRSI framework that can effectively transfer intrusion knowledge for better intrusion detection. 

In summary, the contributions of this paper are four-fold: 
\begin{itemize}
  \item We realise the usefulness to transfer enriched intrusion knowledge from source NI domain to facilitate better intrusion detection for target II domain under an extreme data-scarce unsupervised HDA setting. 
  \item To our best knowledge, we are the pioneer to introduce a recommender system to facilitate better intrusion knowledge transfer. An Adaptive Bi-Recommendation matching mechanism is used to match interests of intrusion categories, forming a positive loop in which bi-recommendation matching and feature space alignment can mutually benefit each other adaptively. The recommender system also benefits hard PL assignment by avoiding PLs that violate mined intrusion interests. 
  \item We utilise a Self-Improving mechanism with four sub-mechanisms, i.e., self-improving hard PL accuracy via voting, self-improving participation and diversity via hybrid PL strategy, self-improving soft PL diversity and confidence and self-improving by exploiting and eliminating error knowledge, and meanwhile prevent error knowledge forgetfulness. 
  \item We conduct comprehensive experiments on five widely recognised intrusion detection datasets to verify the effectiveness of ABRSI, and meanwhile show the usefulness of each constituting component and verify the efficiency. 
\end{itemize}

The rest of the paper is organised as follows: Section \ref{sec:related_work} summarises related works by categories and present our research opportunities, followed by Section \ref{sec:model_preliminary_and_architecture}, which presents model preliminaries and the ABRSI architecture. Section \ref{sec:the_abrsi_algorithm} presents the detailed mechanism constituting the ABRSI model. Experimental setups and result analyses are given in Section \ref{sec:section_experiment}. The last section concludes the paper. We provide an acronym table and a notation table for better readability.

\section{Related Work}\label{sec:related_work}

\textit{Traditional IoT Intrusion Detection} IoT intrusion detection has drawn wide attention from the research community. The rule-based methods perform intrusion detection based on a pre-defined intrusion rule repository. For instance, Chen et al., \cite{jun2014design} proposed complex event processing, which can reactively match events with pre-defined intrusion rules. Later, Dietz et al., \cite{dietz2018iot} proposed to detect intrusions proactively by periodically scanning for pre-defined malicious behaviours among connected IoT devices. As ML techniques evolve quickly, they have been widely adopted to perform intrusion detection. Possible models including isolation forest \cite{eskandari2020passban}, multi-kernel SVM \cite{murali2019lightweight}, and deep learning methods such as capsule network \cite{yao2019capsule}, stacked autoencoder \cite{muhammad2020stacked}, etc. However, these methods either depend on a thorough and up-to-date rule repository, which requires sophisticated expertise to construct with a high cost, or depend on a fully-labelled training dataset, which is time and labour-intensive to annotate, and is especially difficult to acquire due to IoT data scarcity caused by insufficient resource capabilities and associated data privacy issues. Therefore, it provides rooms for domain adaptation-based solutions, which can tackle the data-scarce IoT scenario. 

\textit{\rw{Heterogeneous Domain Adaptation}} Heterogeneous Domain \rw{Adaptation-based} (HDA) methods can transfer knowledge \rw{from a} knowledge-rich source domain \rw{to a} knowledge-scarce target domain, given that heterogeneities \rw{are present} between domains. Hence, it is applicable to facilitate more effective intrusion detection under data-scarce IoT scenarios. Vu et al., \cite{vu2020deep_autoencoder} proposed an autoencoder-based method, which aligned the bottleneck layer to achieve intrusion knowledge transfer. Hu et al., \cite{hu2022deep} presented a deep subdomain adaptation network and transferred intrusion knowledge via local maximum mean discrepancy minimisation. \rw{There are some other HDA methods, although not specifically targeting the intrusion detection scenario, their excellent knowledge transfer capability can be utilised to tackle the intrusion detection problem and achieve excellent performance.} Xie et al., \cite{xie2022collaborative} proposed a collaborative alignment framework (CAF) that performed global domain alignment via Wasserstein distance minimisation (CAFD) or adversarial learning (CAFA). It also achieved knowledge transfer via minimising the distance between probabilistic outputs yielded by two classifiers, a way to learn from errors. Li et al., \cite{li2019joint} considered the JADA model, which jointly achieved both global alignment and local error minimisation between probabilistic outputs of two differently initialised classifiers. However, neither \rw{the} JADA nor CAF method performed probabilistic output matching via a previous knowledge-enabled adversarial way, and their lack of PL information makes their matching process coarse-grained. Li et al., presented DCAN \cite{li2020domain} and GDCAN \cite{li2021generalized}, which transferred knowledge via domain conditional channel attention and the adaptive channel attention mechanism, respectively. However, these methods lacked enough focus on PL assignment, making their knowledge transfer become coarse-grained. On the other hand, Long et al., \cite{long2018conditional} proposed Conditional Domain Adversarial Network (CDAN) to transfer knowledge via a conditional generative adversarial network. It also utilised hard PL assignment generated by the classifier network. Jin et al., \cite{jin2020minimum} put forward a minimum class confusion (MCC) loss as a pluggable tool to generalise and improve the knowledge transfer, assisted by a classifier-predicted hard PL strategy. Lv et al., \cite{liang2021pareto} achieved knowledge transfer via Pareto Domain Adaptation (PDA), a Pareto optimal solution searching method assisted by soft PL assignment. However, these aforementioned methods either assign hard PL in an isolated manner without considering label relationship information, or fail to consider the hybrid PL strategy, resulting in compromised target instance participation and diversity, or a lack of emphasis on confident and correct predictions. \rw{Recently, Li et al., \cite{8475006} presented the progressive alignment framework, which alleviated feature discrepancy via shared codebook learning, and meanwhile minimised the distribution divergence in a progressive manner. This unified framework achieved superior knowledge transfer performance. Li et al., \cite{9528987} further considered a more generalised scenario, i.e., the divergence-agnostic scenario in which either source or target domain data is unknown. They then proposed the AAA model to tackle this scenario via adversarial attacks. }However, all of these methods completely ignored the benefit brought by the recommender system. 

\begin{figure}[!t]
  \begin{center}
    \includegraphics[width=0.48\textwidth,keepaspectratio]{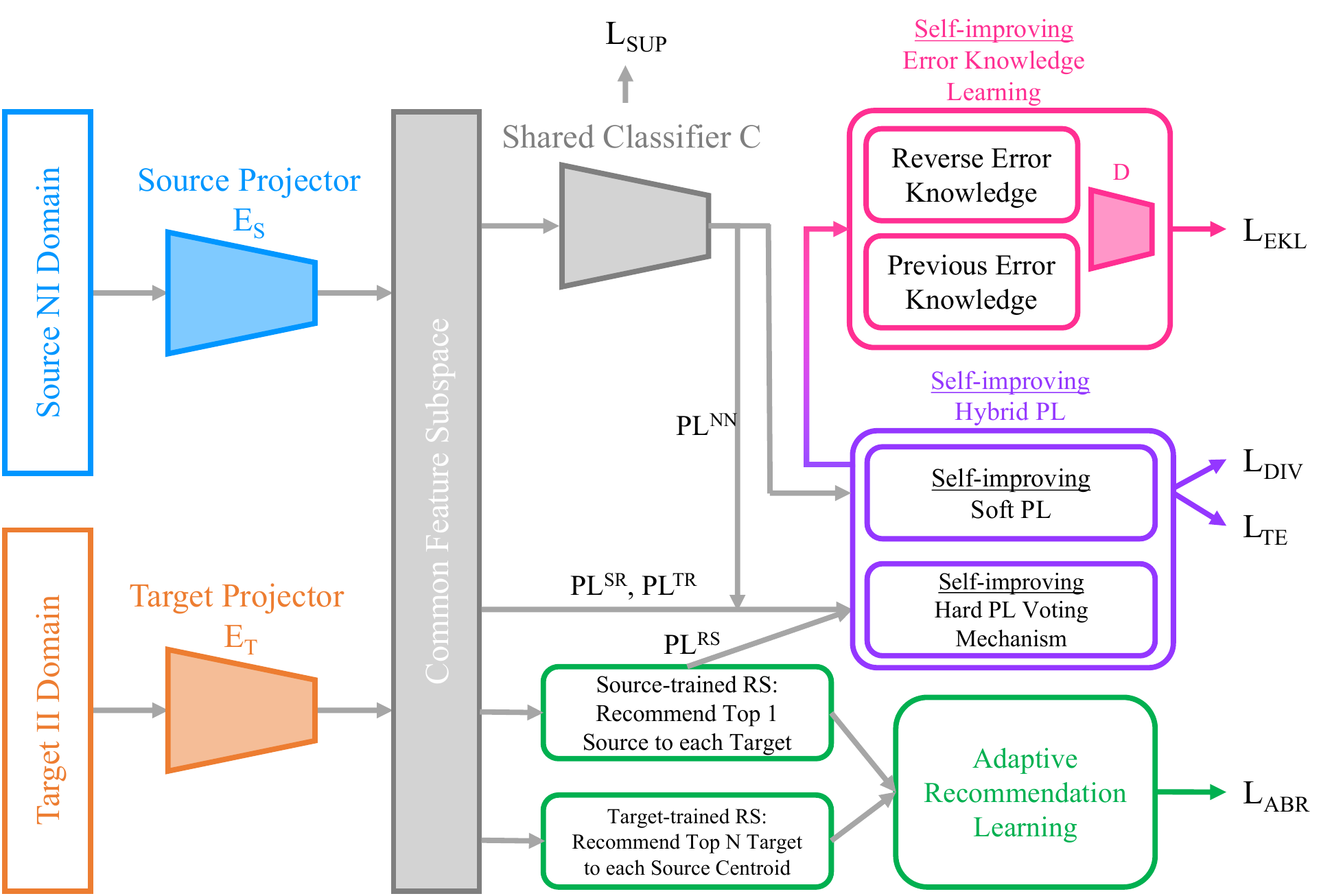}\\
    \caption{The architecture of the ABRSI model}
    \vspace{-0.9cm}
    \label{fig:figure_architecture}
  \end{center}
\end{figure}

\textit{Recommender System and Intrusion Detection} To our best knowledge, we are the pioneer to introduce the recommender system to enhance the intrusion knowledge transfer. To find a comparable baseline, we adapt the transfer learning-based RS algorithm \cite{yuan2019darec} and form ADAR, which utilises domain discriminator to fuse the RS feature matrix, then perform intrusion detection via the trained RS. However, the ADAR does not \rw{involve a} recommendation matching mechanism, and fails to utilise the RS to produce more accurate PLs. 

\textit{Research Opportunity} The ABRSI method jointly utilises the adaptive bi-recommendation mechanism and a self-improving mechanism. The adaptive bi-recommendation mechanism forms a mutually-beneficial positive loop between intrusion feature alignment and interest matching of two recommender systems. The self-improving mechanism benefits fine-grained intrusion knowledge transfer from four ways: it prevents error-prone hard PL with compromised diversity by improving the PL accuracy through a voting mechanism, the RS also contributes during PL voting. Hence, it performs better than the isolated PL strategy used in previous methods; it improves participation and diversity via a hybrid PL strategy, which fills the void of previous methods; it improves the diversity and certainty of soft PL; it improves error knowledge learning and jointly considers previous error knowledge to prevent error forgetfulness, which is lacked by previous methods.

\section{Model Preliminary and Architecture}\label{sec:model_preliminary_and_architecture}

\subsection{Model Preliminary}\label{sec:model_preliminary}

The ABRSI model works under the unsupervised heterogeneous DA setting. We follow common notations \cite{10.1145/3394171.3413995,9933783} to define the source NI domain $\mathcal{D}_{S}$ as follows: 
\begin{equation}
  \begin{split}
    & \mathcal{D}_{S} = \{\mathcal{X}_{S}, \mathcal{Y}_{S}\} = \{(x_{S_{i}}, y_{S_{i}})\}, i \in [1, n_{S}]\,, \\
    & x_{S_{i}} \in \mathbb{R}^{d_{S}}, y_{S_i} \in [1, K]\,, 
  \end{split}
\end{equation}
where $\mathcal{X}_{S}$ is the set of source NI traffic data, $\mathcal{Y}_{S}$ is the corresponding intrusion category label, $n_{S}$ and $d_{S}$ denote the number of source NI instances and dimension of source NI domain, respectively. $K$ denotes the total number of intrusion categories. Similarly, the target II domain has the following definition: 
\begin{equation}
    \mathcal{D}_{T} = \{\mathcal{X}_{T}\} = \{(x_{T_{i}})\}, i \in [1, n_{T}], x_{T_{i}} \in \mathbb{R}^{d_{T}}\,. 
\end{equation}
The ABRSI model works under the unsupervised HDA setting, i.e., the target II domain is completely unlabelled. Besides, heterogeneities present between domains, e.g., $d_S \neq d_T$. 

\subsection{The ABRSI Architecture}\label{sec:abrsi_architecture}

The architecture of the ABRSI model is illustrated in Fig. \ref{fig:figure_architecture}. Each domain has a feature projector that maps heterogeneous features into a $d_C$-dimensional shared feature subspace. The feature projector $E$ has the following definition: 
\begin{equation}
  \begin{split}
    & f(x_i) = \begin{cases}
      E_{S}(x_i) & \text{if $x_i \in \mathcal{X}_S$} \\
      E_{T}(x_i) & \text{if $x_i \in \mathcal{X}_T$}
    \end{cases} \\
    & f(x_i) \in \mathbb{R}^{d_{C}}\,.
  \end{split}
\end{equation}
A shared classifier $C$ is used to make intrusion detection decisions. The ABRSI utilises \rw{the} adaptive bi-recommendation mechanism. It involves a source-trained RS that \rw{recommends the} top one source instance to each target, and a target-trained RS that \rw{recommends the} top N target instances to each source category centroid. Matching the category-wise recommendation between two RSs can enforce finer feature alignment in the shared feature space, and it in turn yields RS \rw{with an} improved recommendation capability. Hence, it forms a positive loop in which bi-recommendation matching and feature alignment mutually benefit each other in an adaptive way. The ABRSI then leverages the self-improving mechanism from four perspectives. A pseudo-labelling voting mechanism is used to improve the accuracy of hard PL assignment by jointly considering the classifier prediction, recommender system decision and label relationship information. It \rw{accounts for} the intrusion interests learned by the RS and avoids error-prone near boundary PLs benefitted from label relationship information. To enhance label diversity and promote better participation of unlabeled target instances during intrusion knowledge transfer, the ABRSI adopts a hybrid PL strategy. The Tsallis entropy and diversity maximisation loss is also used to make the soft PL globally diverse and individually confident. Finally, the ABRSI uses an error knowledge learning mechanism, which exploits error knowledge that causes intrusion detection ambiguities and iteratively eliminates it via adversarially learning with reverse error knowledge and previous error knowledge. The error knowledge learning holistically achieves error knowledge exploitation, elimination, and meanwhile prevents error forgetfulness. Forming these mechanisms into a holistic model achieves fine-grained intrusion knowledge transfer so that the shared classifier $C$ produces the best intrusion detection efficacy for the unlabeled target II domain.

\section{The ABRSI Algorithm}\label{sec:the_abrsi_algorithm}

\subsection{Adaptive Bi-Recommendation (ABR) Matching}\label{sec:adaptive_bi_recommendation_abr}

\begin{figure}[!ht]
  \begin{center}
    \includegraphics[width=0.48\textwidth,keepaspectratio]{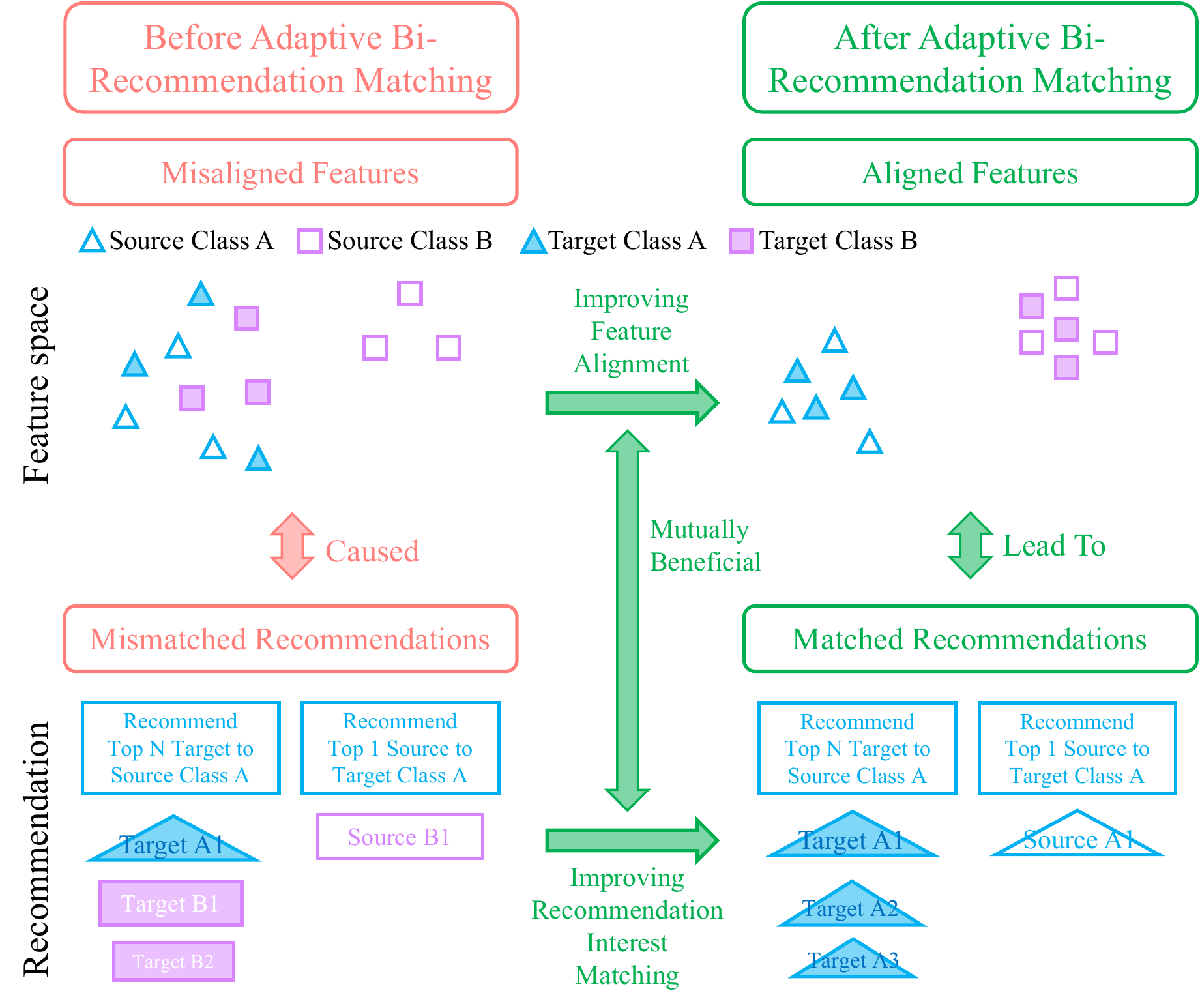}\\
    \caption{Adaptive bi-recommendation matching and feature alignment mutually benefit each other, forming a positive loop that enhances fine-grained intrusion knowledge transfer}
    \vspace{-0.5cm}
    \label{fig:figure_recommendation}
  \end{center}
\end{figure}

Recommender system (RS) is a powerful tool to exploit interests, i.e., characteristics of each intrusion category. However, the RS for each domain cannot find similar interests without the features being well-aligned as illustrated in Fig. \ref{fig:figure_recommendation} (left). The adaptive bi-recommendation bridges the gap between bi-recommendation matching and fine-grained intrusion domain alignment. The ABRSI utilises a source-trained RS, denoted as $RS_{S}$. Since both domains are mapped into a common feature space, therefore, the source-trained RS should be able to recommend the most similar source intrusion instance to a given target instance. If the RS fully exploits intrusion category interests, then, recommending an intrusion type B source instance to a given unlabelled target instance indicates that target instance is highly likely to belong to intrusion type B\rw{, i.e., the top one recommendation for each unlabelled target instance forms a RS-based pseudo label $\text{PL}_{\text{RS}}$. }\rw{Note that in this case, only the top one recommendation is used since it needs to act as the pseudo label for unlabelled target instances. }By averaging target instances based on its RS-based PL, it forms the intrusion category-wise RS recommendation for the target domain. Similarly, a target-trained $RS_{T}$ is utilised to recommend top N most similar target instances to each source category centroid, and take the average of these N recommendations as a final recommendation for each source intrusion category. \rw{Note that in this case the top N recommended target instances are used for a better representativeness of target domain interests.} By enforcing the category-wise recommendation to be similar between two RSs, it in turn forces the intrusion domains to align in the shared feature space, as illustrated in Fig. \ref{fig:figure_recommendation} since misaligned feature space can only cause mismatched RS recommendations. Moreover, the aligned features can also facilitate the bi-RS to better exploit and match the intrusion category interests, hence, it forms a positive loop in which bi-recommendation interest matching and feature alignment mutually benefit each other as the training progresses. As shown in Fig. \ref{fig:figure_recommendation} (right), the ABR mechanism will eventually achieve a well-aligned recommendation interest matching and a fine-grained feature space alignment.

\begin{figure*}[!ht]
  \begin{center}
    \includegraphics[width=\textwidth,keepaspectratio]{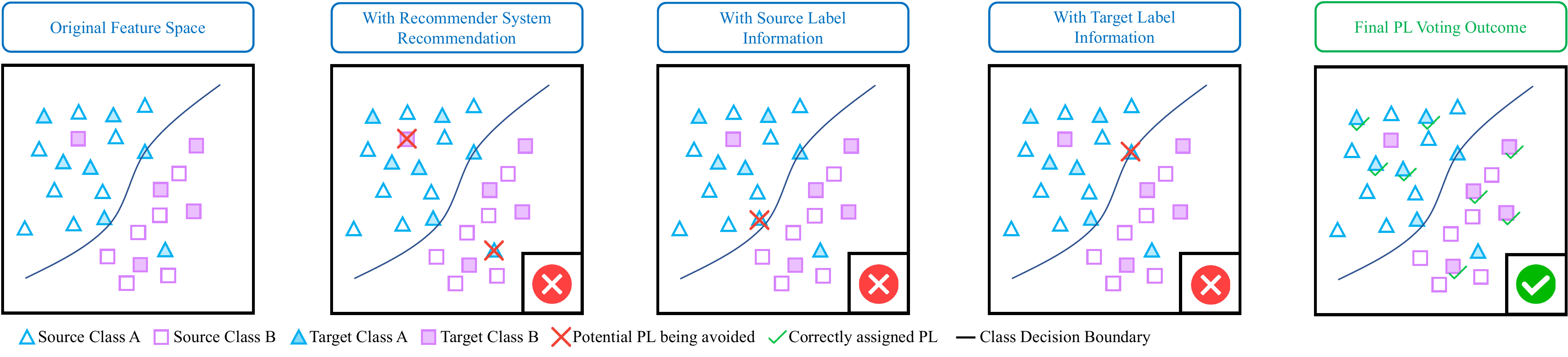}\\
    \caption{Pseudo label voting mechanism. (a) the original feature space; (b-d) PLs eliminated by RS recommendation, source and target label relationship information, respectively; (e) the hard PL assigned by pseudo label voting mechanism}
    \vspace{-0.5cm}
    \label{fig:figure_pseudo_label}
  \end{center}
\end{figure*}

{\renewcommand{\arraystretch}{1.2}%
\begin{table*}[!ht]
  \caption{Hybrid PL strategy and its advantages over hard and soft-PL-only strategies}
  \centering
  \label{tab:pseudo_labelling_strategy}
  \begin{tabular}{c|p{3cm}|p{3cm}|p{8cm}}
  \Xhline{2\arrayrulewidth}
  Strategy & \centering Use Hard PL & \centering Use Soft PL & Pros \& Cons \\ \hline
  Hard PL only & \centering\checkmark & \centering\xmark & \checkmark\hspace{1mm}Emphasis on confident and correct PLs\newline\xmark\hspace{1mm}Error prone and misleading\newline\xmark\hspace{1mm}Instances without hard PL cannot participate during alignment\newline\xmark\hspace{1mm}Hard PLs favour easy instances, lacking diversity\\ \hline
  Soft PL only & \centering\xmark & \centering\checkmark & \checkmark\hspace{1mm}All instances can participate during alignment\newline\checkmark\hspace{1mm}Promotes better diversity using probabilistic output as PL\newline\xmark\hspace{1mm}Confident and correct predictions get less emphasis unlike hard PL\\ \hline
  \rowcolor{gray}
  ABRSI Hybrid PL & \centering\checkmark\hspace{1mm}Hard PL with\newline PL Voting Mechanism & \centering\checkmark\hspace{1mm}Soft PL with\newline $\mathcal{L}_{TE}$ and $\mathcal{L}_{DIV}$ & \checkmark\hspace{1mm}PL Voting Mechanism increases hard PL correctness\newline\checkmark\hspace{1mm}Hard PL emphasises confident and correct predictions\newline\checkmark\hspace{1mm}Soft PL boosts target instance participation during alignment\newline\checkmark\hspace{1mm}$\mathcal{L}_{DIV}$ improves soft PL global  diversity\newline\checkmark\hspace{1mm}$\mathcal{L}_{TE}$ gradually increases individual soft PL certainty\\ \Xhline{2\arrayrulewidth}
  \end{tabular}
  \vspace{-3mm}
\end{table*}}

The ABRSI model utilises Latent Semantic Indexing (LSI) algorithm as its RS. \rw{Unlike neural network-based RS algorithms that involve a tremendous amount of parameters \cite{nfmpaper}, the LSI is not heavily data-hungry during training unlike its neural-network-based counterparts, and enjoys a low computational complexity during intrusion detection inference. Besides, it has been leveraged in various recommendation system applications \cite{10.1145/3172944.3172979,lsiapp} and achieved} a satisfying recommendation effectiveness, which makes it suitable to be used under data-scarce and computationally-constrained IoT scenarios. The formulation of LSI is as follows: 
\begin{equation}
  \begin{split}
    & M_{S} \approx U_{S}^{R}T_{S}^{R}V^{R\top}_{S}, U_{S}^{R} = \begin{bmatrix}
      x_{S}^{1'} \\
      x_{S}^{2'} \\
      \vdots \\
      x_{S}^{n_{S}'} \\
    \end{bmatrix}, x_{T}^{j'} = x_{T}^{j}U_{S}^{R}T^{R\top}_{S}\\
    & M_{S} \in \mathbb{R}^{d_{C} \times n_{S}}, U_{S}^{R} \in \mathbb{R}^{d_{C} \times R}, T_{S}^{R} \in \mathbb{R}^{R \times R}, V^{R\top}_{S} \in \mathbb{R}^{n_{S} \times R}, \\
    & x_{S}^{i}, x_{T}^{j} \in \mathbb{R}^{d_{C}}, x_{S}^{i'}, x_{T}^{j'} \in \mathbb{R}^{R}\,, \\
  \end{split}
\end{equation}
where $M_{S}$ is the original feature matrix of source NI domain, $R$ is the reduced dimension parameter, $U^{R}_{S}$, $T^{R}_{S}$ and $V^{R\top}_{S}$ are the feature-latent, latent transfer and instance-latent matrix for source domain $D_S$, respectively. $x_{S}^{i'}$ denotes the transformed $i$\textsuperscript{th} source NI instance. Then, the top one recommendation for each target II instance $x_{T}^{j}$ is made as follows: 
\begin{equation}
  RS_S(x_{T}^{j}) = x_{S}^i, PL^{RS}_{x_{T}^{j}} = y_{S}^i, i = \argmax_i COS(x_{S}^{i'}, x_{T}^{j'})
\end{equation}
The target-trained $RS_T$ is defined similarly with the above definition of $RS_S$. The $RS_S$ will recommend the most similar source instance $RS_S(x_{T}^{j})$ to each target instance $x_{T}^{j}$, its category label acts as the RS label $PL^{RS}_{x_{T}^{j}}$, and the $RS_T$ will recommend the top N most similar target instances $RS_T(x_{S}^{(i)})^n, n \in [1, N]$ to each source category centroid $x_{S}^{(i)}$. The ABR will then minimise the divergence between category-wise bi-recommendations as follows: 
\begin{equation}
  \begin{split}
    \mu_{RS_S}^k &= \text{mean}(RS_S(x_{T}^{j})), \text{where } PL^{RS}_{x_{T}^{j}} = k, \\
    \mu_{RS_T}^k &= \sum_{n=1}^{N}\frac{(RS_T(x_{S}^i)^n)}{N}, \text{where } y_S^i = k, \\
    \mathcal{L}_{ABR} &= \sum_{k=1}^{K}\frac{||\mu_{RS_S}^k - \mu_{RS_T}^k||_2^2}{K}\,,
  \end{split}
\end{equation}
where $\mu_{RS_S}^k$ denotes the centroid of target instances that are recommended with a category $k$ source instance, $\mu_{RS_T}^k$ denotes the centroid of target instances recommended to the centroid of source category $k$. As the training progresses, the recommendations produced by bi-RSs will become more reliable, and hence the adaptive bi-recommendation matching is gradually emphasised by a linearly-growing hyperparameter $\rho$. The adaptive bi-recommendation loss $\mathcal{L}_{ABR}$ minimises the category-wise divergence between bi-recommendations, which therefore promotes better feature space alignment, forming a positive loop as indicated in Fig. \ref{fig:figure_recommendation}.

\subsection{Self-Improving Mechanism (SI)}\label{sec:self_improving_mechanism_si}

\textbf{Hard Pseudo Label Voting} To give unlabelled target instances a chance to participate during fine-grained intrusion knowledge transfer, assigning hard pseudo labels to them is a common way to fully exploit their potentials. However, previous methods tend to directly use the error-prone network predicted PL, which misleads the intrusion knowledge transfer and causes negative transfer. To mitigate their drawbacks, the ABRSI uses a hard pseudo label voting mechanism. Firstly, the ABRSI will use the recommender system decision $PL^{RS}_{x_{T}^{j}}$ introduced earlier to avoid confident but wrong network predicted PLs that violates the interests mined by the RS, as indicated in Fig. \ref{fig:figure_pseudo_label}.2. We are the pioneer to use a recommender system to improve hard PL accuracy for better intrusion knowledge transfer. Then, the ABRSI will consider the PL relationship with surrounding source instances. If the neighbouring source instances cannot reach a consensus, or the consensus contradicts with the network predicted PL or the RS decided PL, then that PL is highly likely to reside near the decision boundary, i.e., ambiguous near-boundary PL, as indicated in Fig. \ref{fig:figure_pseudo_label}.3. Hence, the ABRSI will avoid this assignment when the source label relationship information yields disagreements. Finally, the ABRSI performs unsupervised clustering on unlabelled target instances to reveal the target label relationship. If a target instance has a PL assignment that contradicts with its within-cluster peers, then that instance is also considered as an ambiguous near-boundary instance, and will be eliminated as shown in Fig. \ref{fig:figure_pseudo_label}.4. Eventually, the ABRSI will only assign a hard PL to an instance if four votes reach a consensus. As shown in Fig. \ref{fig:figure_pseudo_label}.5, the assigned \rw{hard PLs} after PL voting are highly accurate. Hard PL assignment with improved accuracy can therefore positively guide the intrusion knowledge transfer. 

\textbf{Hybrid Pseudo Label Strategy} Previous research efforts either \rw{used a} hard-PL-only strategy that \rw{lacks the} recommender system decision and label relationship information, or used a probabilistic soft-PL-only strategy. As summarised in Table. \ref{tab:pseudo_labelling_strategy}, the hard-PL-only strategy suffers from negative transfer caused by its erroneous assignment. Moreover, instances without hard PL agreement cannot \rw{participate in the} intrusion knowledge transfer. Besides, hard PL favours easy-to-predict categories with impaired diversity. Conversely, the soft-PL-only strategy promotes better participation and diversity, \rw{but has} insufficient emphasis on confident and correct predictions. If these predictions get more emphasis as in hard PL, they can positively contribute towards finer intrusion alignment. 

To tackle the deficiencies summarised in Table. \ref{tab:pseudo_labelling_strategy}, the ABRSI leverages a hybrid PL strategy. Instances that have four voter agreements will be assigned with a hard PL to promote positive contributions of correct and confident predictions. The remaining instances will have a probabilistic soft PL assignment to promote full participation. The soft PL also enriches diversity, which tackles the impaired diversity caused by hard PL's potential focus on easy intrusion categories. Finally, we use a diversity maximisation loss $\mathcal{L}_{DIV}$ to improve soft PL global diversity, which is defined as follows: 
\begin{equation}
  \begin{split}
    \mathcal{L}_{DIV} = \sum_{k=1}^{K} {p_{T}^{\mu}}^{(k)} \text{log} {p_{T}^{\mu}}^{(k)}, p_{T}^{\mu} = \frac{1}{n_{T}}\sum_{j=1}^{n_T} p_{T}^j\,,
  \end{split}
\end{equation}
where $p_{T}^j$ stands for the probabilistic output of target instance $x_T^j$ from the shared classifier $C$, $p_{T}^{\mu}$ is the mean of all target probabilistic outputs, ${p_{T}^{\mu}}^{(k)}$ is the $k$\textsuperscript{th} element of $p_{T}^{\mu}$. Minimising the $\mathcal{L}_{DIV}$ promotes better global soft PL diversity. 

Meanwhile, the ideal soft PL should be globally diverse while individually certain. Hence, we use $\alpha$\textit{-Tsallis Entropy} $\mathcal{L}_{TE}$ to promote individual soft PL certainty as follows: 
\begin{equation}
  \mathcal{L}_{TE} = \sum_{j = 1}^{n_T} (\frac{1}{\alpha - 1} (1 - \sum_{k} {{p_{T}^{j}}^{(k)}}^{\alpha})), \alpha > 0 \,,
\end{equation}
where ${p_{T}^{j}}^{(k)}$ denotes the $k$\textsuperscript{th} element of the probabilistic output of target instance $x_T^j$. \rw{The $\alpha$\textit{-Tsallis entropy} applies an adjustable entropic index $\alpha$.} A higher $\alpha$ value \rw{places a lower} penalty on uncertain predictions, and vice versa. \rw{Hence, the ABRSI linearly decreases $\alpha$ between $\alpha_{max}$ and $\alpha_{min}$ to adaptively promote certainty for soft PLs as the training progresses. }

By combining a hard PL strategy with improved accuracy, a soft PL strategy with global diversity and individual certainty, the ABRSI forms a hybrid PL strategy with improved diversity, participation, and enough emphasis on confident and correct predictions. The improved PL strategy can therefore benefit finer intrusion knowledge transfer.

\begin{figure}[!ht]
  \begin{center}
    \includegraphics[width=0.49\textwidth,keepaspectratio]{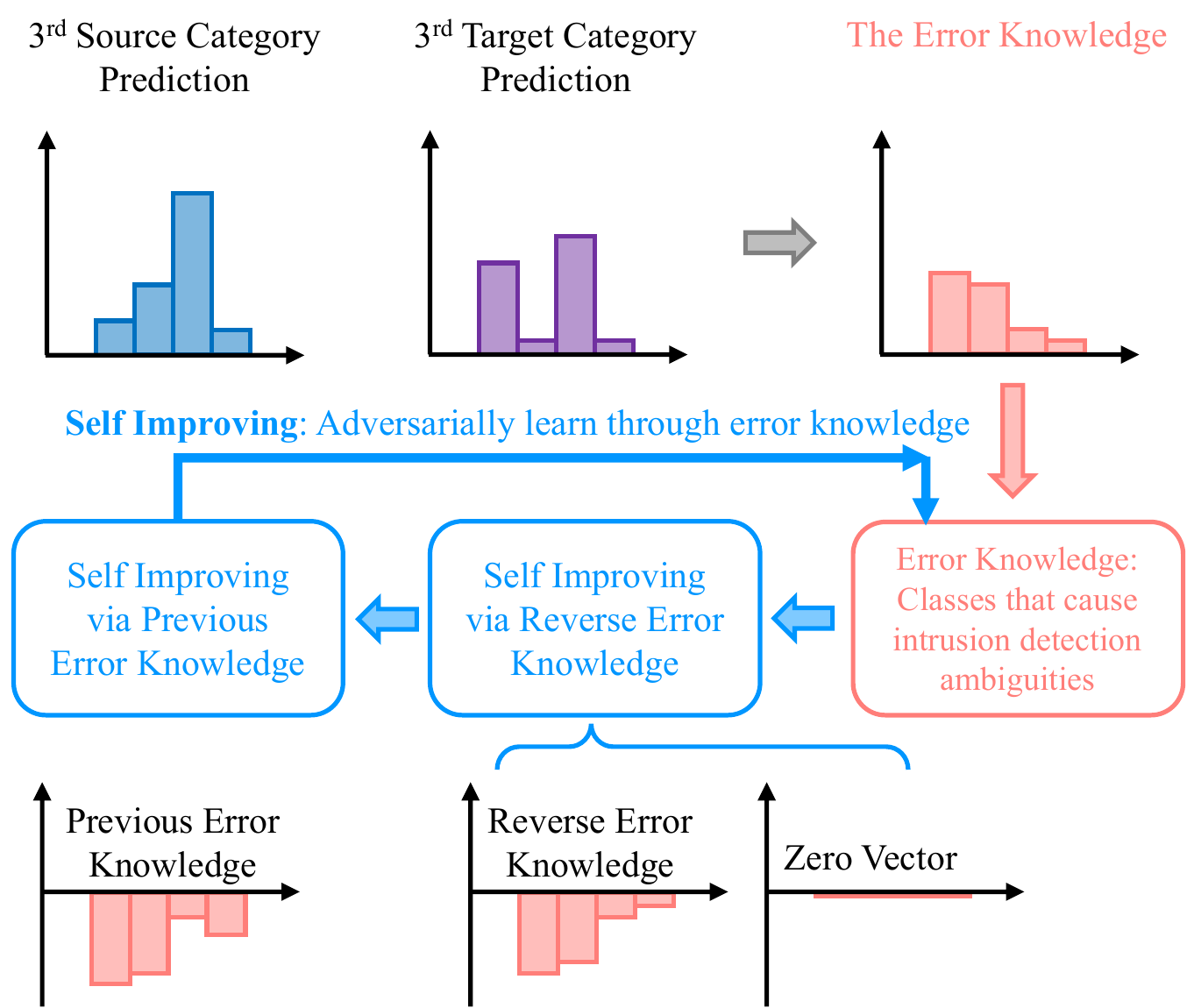}\\
    \caption{Illustration of self-improving via error knowledge learning, the $3$\textsuperscript{rd} category is chosen as an example}
    \vspace{-0.5cm}
    \label{fig:figure_error_knowledge}
  \end{center}
\end{figure}

\textbf{Error Knowledge Learning (EKL)} Degraded intrusion knowledge transfer is stemmed from error factors that cause ambiguities. For example, as illustrated in Fig. \ref{fig:figure_error_knowledge}, any probabilistic output other than the third category are factors that causes ambiguities and can potentially lead to compromised intrusion detection performance. To eliminate these error factors, the ABRSI model forms it as a learning process. Firstly, the ABRSI model extracts the error knowledge by finding the category-wise divergence between probabilistic outputs between domains. Specifically, the probabilistic output \rw{of the} $k$\textsuperscript{th} source intrusion category ${p_{S}^{\mu}}^{(k)}$ is defined as follows: 
\begin{equation}
  \begin{split}
    {p_{S}^{\mu}}^{(k)} &= \frac{1}{n_S} \sum_{i = 1}^{n_S} p_{S}^i, \text{where } y_{S}^i = k\,,
  \end{split}
\end{equation}
and the probabilistic output of the $k$\textsuperscript{th} target intrusion category ${p_{T}^{\mu}}^{(k)}$ is defined as follows: 
\begin{equation}
  \begin{split}
    {p_{T}^{\mu}}^{(k)} &= \frac{1}{\sum_{j = 1}^{n_T} {p_{T}^j}^{(k)}} \sum_{i = 1}^{n_T} ({p_{T}^j}^{(k)} \times p_{T}^j), \\
    p_{T}^j &= \begin{cases}
      \mathds{1}(PL^{NN}_{x_{T}^{j}}), \text{if } PL^{NN}_{x_{T}^{j}}\text{=}PL^{RS}_{x_{T}^{j}}\text{=}PL^{SR}_{x_{T}^{j}}\text{=}PL^{TR}_{x_{T}^{j}} \\
      C(f(x_T^j)), \text{otherwise}
    \end{cases}
  \end{split}
\end{equation}
where $p_{T}^j$ is the one hot PL vector if a hard PL is assigned to $x_T^j$, denoted by $\mathds{1}(PL^{NN}_{x_{T}^{j}})$, or otherwise it is the probabilistic output yielded by the shared classifier $C$. ${p_{T}^j}^{(k)}$ denotes the $k$\textsuperscript{th} element of $p_{T}^j$, $PL^{NN}_{x_{T}^{j}}, PL^{RS}_{x_{T}^{j}}, PL^{SR}_{x_{T}^{j}}, PL^{TR}_{x_{T}^{j}}$ denote the neural network predicted PL, recommender system decided PL, source label relationship PL and target label relationship PL, respectively. The ABRSI model then gets the category $k$ error knowledge between probabilistic outputs, i.e., $EK^{(k)}$, defined as follows: 
\begin{equation}
  EK^{(k)} = ||{p_{S}^{\mu}}^{(k)} - {p_{T}^{\mu}}^{(k)}||_2^2\,.
\end{equation}

Intuitively, feeding the category-wise error knowledge vector $EK^{(k)}$ and the zero vector $EK_{0}$ into a discriminator D would let the discriminator \rw{try to} distinguish them as much as possible, i.e., exploiting the error knowledge as much as possible, and simultaneously let the feature projector $E$ and shared classifier $C$ to confuse the discriminator to their best extent, i.e., learn to eliminate the error knowledge. \rw{Once this} minimax game reaches an equilibrium, the error knowledge can be fully exploited and then eliminated at the category level. However, we find out that merely using the zero vector $EK_{0}$ may not produce an error correction signal that is strong enough for the error knowledge learning. Therefore, we further apply a reverse error knowledge vector $EK^{(k)}_{\mathcal{R}}$, which stretches the divergence slightly towards the reverse direction by multiplying the error knowledge vector $EK^{(k)}_{\mathcal{R}}$ with a negative constant $\psi$. 

We further realise that correcting the error knowledge in the current round may damage the error correction learned previously, i.e., error knowledge forgetfulness. For instance, eliminating some error factors may cause some other error factors to emerge again. Therefore, to further reinforce the error learning efficacy, we also feed the error knowledge of the previous round $\phi EK_{\mathcal{P}}^{(k)}$ to prevent error knowledge forgetfulness from happening. $\phi$ is a hyperparameter that balances current error knowledge learning and error knowledge forgetfulness prevention. Overall, the error knowledge learning loss is defined as follows: 
\begin{equation}
  \begin{split}
    &\mathcal{L}_{EKL} = \frac{1}{K}\sum_{k = 1}^{K}\text{log}(D(EK^{(k)})) + \frac{1}{3K}\sum_{k=1}^{K}(3-\text{log}(D(EK_{*}^{(k)})),\\
    &EK_{*}^{(k)} \in [EK_{0}, EK_{\mathcal{R}}^{(k)}, EK_{\mathcal{P}}^{(k)}]\,.
  \end{split}
\end{equation}
By letting the feature projector and classifier to confuse the discriminator with $EK^{(K)}$, $EK_{0}$, $EK_{\mathcal{R}}$ and $EK_{\mathcal{P}}$ and carry out this minimax game in an adversarial manner, the error knowledge learning process forms an error knowledge extraction and elimination cycle with error knowledge forgetfulness prevention, which will gradually improve the intrusion knowledge transfer, as illustrated in Fig. \ref{fig:figure_error_knowledge}. 

\subsection{Overall Optimisation Objective}\label{sec:overall_optimisation_objective}

Finally, the source label can provide supervision during intrusion knowledge alignment, which is defined as follows: 
\begin{equation}
  \mathcal{L}_{SUP} = \frac{1}{n_S} \sum_{i = 1}^{n_S} \mathcal{L}_{CE}(C(f(x_S^i)), y_S^i)\,,
\end{equation}
where $\mathcal{L}_{CE}$ denotes cross entropy loss. The overall optimisation objective of the ABRSI model is as follows: 
\begin{equation}
  \begin{split}
    &\min_{E_S, E_T, C}(\mathcal{L}_{SUP} + \rho \mathcal{L}_{ABR} + \delta \mathcal{L}_{DIV} + \tau \mathcal{L}_{TE} + \gamma \mathcal{L}_{EKL}), \\
    &\max_{D}(\gamma \mathcal{L}_{EKL})
  \end{split}
\end{equation}
where $\rho$, $\gamma$, $\delta$ and $\tau$ are hyperparameters controlling the weight of the corresponding components. To achieve an end-to-end optimisation process, a gradient reversal layer \cite{ganin2016domain} is added for the discriminator, which performs the role \rw{of an} identity function during forward propagation, and reverses the gradient during backpropagation. \rw{Once the} above minimax game reaches an equilibrium, the intrusion knowledge can be transferred in a fine-grained manner, so that the shared classifier can perform intrusion detection as accurate as possible.


\section{Experiment}\label{sec:section_experiment}

We utilise five comprehensive and representative intrusion detection datasets and nine state-of-the-art comparing baseline methods to verify the effectiveness of the ABRSI model. In addition, the contribution and necessity of each ABRSI constituting component, and the computational efficiency of the ABRSI model is verified. 

\subsection{Experimental Datasets}\label{sec:experimental_datasets}

We use five comprehensive intrusion detection datasets, including three network intrusion datasets: NSL-KDD, UNSW-NB15 and CICIDS2017, and two IoT intrusion detection datasets: UNSW-BOTIOT and UNSW-TONIOT. 

\textbf{Network Intrusion Dataset: NSL-KDD} Released in 2009, the NSL-KDD dataset \cite{tavallaee2009nslkdd} has a better data quality than its previous version \cite{hettich1999uci}. The dataset contains benign traffic with four types of representative intrusions, such as denial of service attack, probing attack, etc. We utilise $20\%$ of the dataset, a reasonable amount following other works such as \cite{anthi2019smart_home_iot}. The dataset adopts a 41-dimensional feature representation. We follow Harb \cite{harb2011selecting} to use the top-31 most informative features as the feature space. The dataset is denoted as $K$.

{\renewcommand{\arraystretch}{1.2}%
\begin{table*}[!ht]
  \caption{Intrusion detection accuracy of $10$ methods on nine tasks}
  \vspace{-0.3cm}
  \centering
  {\renewcommand{\arraystretch}{1.3}
  \begin{tabular}{c|ccccccccc|c}
  \Xhline{2\arrayrulewidth}
  $\mathcal{D}_{S} \rightarrow \mathcal{D}_{T}$ & N $\rightarrow$ B$^*$ & N $\rightarrow$ G & N $\rightarrow$ W & N $\rightarrow$ M & C $\rightarrow$ G & C $\rightarrow$ B & K $\rightarrow$ B$^*$ & K $\rightarrow$ G$^*$ & K $\rightarrow$ W & Avg \\ \hline
  CDAN & $40.17$ & $49.89$ & $52.90$ & $50.23$ & $49.75$ & $49.65$ & $34.22$ & $24.84$ & $53.78$ & $45.05$ \\
  MCC & $35.96$ & $50.02$ & $50.18$ & $51.20$ & $49.85$ & $57.88$ & $39.66$ & $24.86$ & $53.55$ & $45.61$ \\
  CAFD & $33.60$ & $49.72$ & $55.51$ & $49.90$ & $50.10$ & $60.34$ & $36.38$ & $24.80$ & $50.30$ & $45.63$ \\ 
  JADA & $32.20$ & $50.10$ & $51.40$ & $51.55$ & $50.80$ & $57.85$ & $38.52$ & $24.80$ & $56.04$ & $45.92$ \\ 
  CAFA & $36.45$ & $49.15$ & $53.45$ & $51.28$ & $50.88$ & $62.34$ & $38.32$ & $24.85$ & $50.43$ & $46.35$ \\
  DCAN & $34.97$ & $51.00$ & $49.32$ & $51.25$ & $50.30$ & $59.38$ & $40.20$ & $24.90$ & $55.85$ & $46.35$ \\
  GDCAN & $37.23$ & $49.75$ & $51.03$ & $51.68$ & $50.20$ & $60.26$ & $38.43$ & $24.85$ & $56.50$ & $46.66$ \\
  PDA & $40.06$ & $71.88$ & $59.38$ & $52.43$ & $60.66$ & $54.44$ & $43.84$ & $31.25$ & $55.79$ & $52.19$ \\
  ADAR & $45.75$ & $70.66$ & $55.92$ & $51.54$ & $63.21$ & $58.00$ & $42.04$ & $39.73$ & $57.40$ & $53.81$ \\ \hline
  \rowcolor{gray}
  \textbf{ABRSI (Ours)} & $\textbf{46.93}$ & $\textbf{90.31}$ & $\textbf{67.45}$ & $\textbf{53.17}$ & $\textbf{88.04}$ & $\textbf{65.42}$ & $\textbf{47.42}$ & $\textbf{47.27}$ & $\textbf{61.22}$ & $\textbf{63.03}$ \\ \Xhline{2\arrayrulewidth}
  \end{tabular}}
  \label{tab:giant_performance_table}
  \vspace{-0.3cm}
\end{table*}}

\begin{figure*}[t]
  \begin{center}
    \includegraphics[width=\textwidth,keepaspectratio]{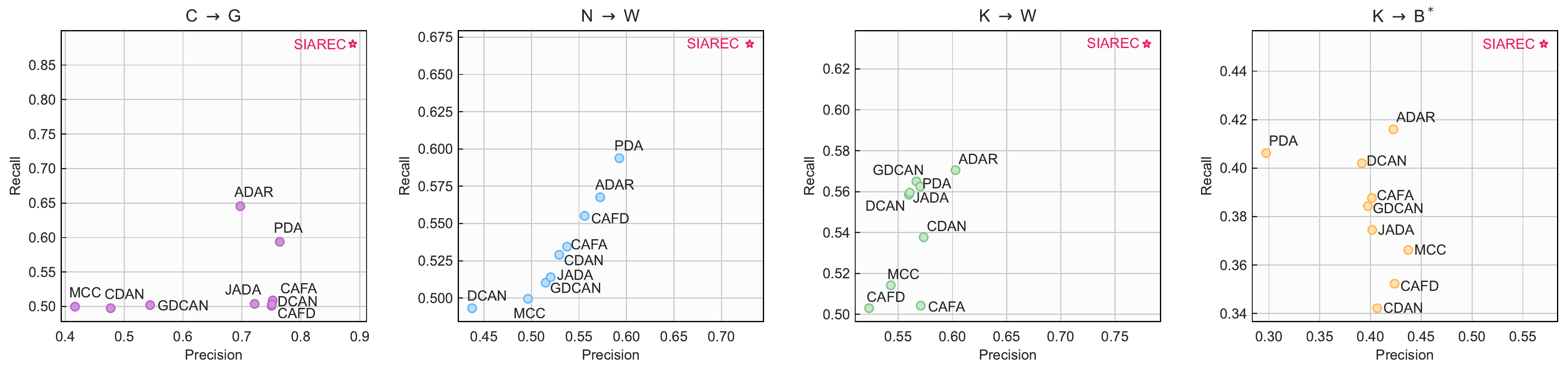}\\
    \caption{Intrusion detection performance evaluation using precision (X-axis) and recall (Y-axis) of $10$ methods on four randomly selected tasks}
    \vspace{-0.5cm}
    \label{fig:figure_metrics_pr}
  \end{center}
\end{figure*}

\textbf{Network Intrusion Dataset: UNSW-NB15} The UNSW released this dataset \cite{moustafa2015unsw} in 2015, captured using the Ixia PerfectStorm tool, a comprehensive security test platform commonly used by the industry. The dataset has high quality and up-to-date comprehensive network flows. It contains normal network traffic and nine categories of representative intrusions, such as DoS attacks, reconnaissance attacks, etc. Following the dataset magnitude in \cite{alkadi2020deep}, we utilise 6500 traffic records. We also perform preprocessing to reduce the feature dimension from 49 to 45 by removing four features having a value zero for nearly all records. The dataset is denoted as $N$. 

\textbf{Network Intrusion Dataset: CICIDS2017} As one of the most up-to-date network intrusion datasets, this dataset \cite{sharafaldin2018toward} was released in 2017. The dataset captures seven intrusion types, represented in 77 dimensions. We utilise $20\%$ of the dataset provided by its creator. The preprocessing steps including data deduplication and categorical-numerical data conversion. Guided by Stiawan \cite{stiawan2020cicids}, we use features with the top 40 information gain as the feature representation. The dataset is denoted as $C$. 

\textbf{IoT Intrusion Dataset: UNSW-BOTIOT} This dataset \cite{koroniotis2019towards} was released in 2017. It contains up-to-date modern attack models, tested on a realistic testbed that contains commonly-used IoT devices such as weather station, smart fridge, etc. The testbed also uses the MQTT protocol, a lightweight communication protocol commonly used by IoT. The dataset has four categories of modern intrusion types, including information theft attacks, DoS attacks, etc. We follow \cite{alkadi2020deep} to use around 10,000 data records. We follow dataset creator's advice to use the top 10 most informative features among 46 original features and denote the dataset as $B$. 

\textbf{IoT Intrusion Dataset: UNSW-TONIOT} The UNSW-TONIOT \cite{booij2021ton_iot} dataset is one of the most up-to-date IoT intrusion datasets, released in 2021. It involves the latest IoT standards, protocols and uses a sophisticated testbed, consisting of seven types of real IoT devices, including modbus sensors, weather meters, GPS trackers, etc. The dataset contains nine types of common IoT intrusions \cite{abdelmoumin2021performance}, such as scanning attacks, DoS attacks, etc. Each type of IoT device captures data in its own dimension, causing heterogeneities between data domains. Following \cite{qiu2020adversarial}, we utilise around $10\%$ of the dataset, and select three IoT devices, i.e., weather meter, GPS tracker and modbus sensor, denoted as $W$, $G$ and $M$, respectively. 

\textbf{Dataset Comprehensiveness and Intrusion Methods} The datasets we used are comprehensive and representative, making them sufficient to verify the proposed model. Firstly, these datasets are widely used and recognised by the intrusion detection research community with thousands of citations. Secondly, all datasets are developed and released in recent years, some of them are released in 2021. Hence, these datasets reflect modern attack methods. Thirdly, these datasets are captured using widely recognised testbeds with comprehensiveness. The IoT datasets also involve real-world IoT devices, deployed in realistic environments. Finally, the network and IoT datasets have at most eight shared intrusion categories transferrable as intrusion knowledge. \rw{These intrusions are generated using various different mechanisms to reflect the diversity of modern intrusions, and have a wide coverage of modern intrusion trends, i.e., with} a $100\%$, $55\%$, $100\%$, $100\%$ and $98\%$ coverage on NSL-KDD, UNSW-NB15, CICIDS2017, UNSW-BOTIOT and UNSW-TONIOT datasets, respectively. Together, these rationales make these datasets sufficient to testify the effectiveness of the proposed method. 

\subsection{Implementation Details}\label{sec:implementation_details}

We use \rw{the} deep learning framework PyTorch to implement the ABRSI model. We follow \cite{10.1145/3394171.3413995,9933783} to implement the feature projectors as two-layer neural networks with LeakyRelu activation function, and implement both the shared classifier $C$ and EKL discriminator $D$ as single-layer neural networks. All experiments use a single set of hyperparameters, which are as follows: $\rho_{max} = 0.1$, $\delta = 1$, $\tau = 0.005$, $\gamma = 0.1$, $TopN = 3$, \#$\text{PL}^{\text{SR}}\ neighbour = 3$, $\alpha_{max} = 8$, $\alpha_{min} = 4$, $\psi = -0.3$ and $\phi = -0.05$. Note that the hyperparameter $\rho$ will linearly increase from $0$ to $\rho_{max}$ to gradually emphasise the importance of adaptive bi-recommendation matching as the recommendation becomes more mature as training progresses. Besides, hyperparameter $\alpha$ will linearly decrease from $\alpha_{max}$ to $\alpha_{min}$ to gradually increase the soft PL individual certainty as the training evolves. We also perform parameter sensitivity analysis to verify the stability and robustness of ABRSI on hyperparameter settings. During our experiments, we consider both binary intrusion detection, i.e., distinguish benign traffic and intrusion traffic as a whole, and multi-class intrusion detection, i.e., distinguish benign traffic and each individual intrusion type. We denote multi-class intrusion detection with superscript (*) on the target domain. Following \cite{alkadi2020deep,li2018ai_accuracy} and \cite{9933783}, we utilise unlabelled target domain prediction accuracy, category-weighted precision (P), recall (R), F1-Score (F) and Area under the ROC Curve (A) as the evaluation metrics. 

\subsection{State-of-the-art Baselines}\label{sec:state_of_the_art_baselines}

We use nine state-of-the-art baselines to verify the superiority of the ABRSI model, which include CDAN \cite{long2018conditional}, MCC \cite{jin2020minimum}, CAFA \cite{xie2022collaborative}, CAFD \cite{xie2022collaborative}, JADA \cite{li2019joint}, DCAN \cite{li2020domain}, GDCAN \cite{li2021generalized}, PDA \cite{liang2021pareto} and ADAR \cite{yuan2019darec}. All methods are from top-tier venues, and eight of them are published between 2019 and 2022. We summarise their differences with ABRSI as follows: 
\begin{itemize}
  \item From the recommender system usage perspective, the ADAR utilises domain discriminator to facilitate the learning of a recommender system, and lets it perform intrusion category recommendation. However, the ADAR uses a single RS and lacks adaptive bi-recommendation matching. It also fails to use RS to facilitate better pseudo label prediction. 
  \item From the pseudo label perspective, methods such as CDAN utilise pure hard PL, while methods such as MCC and PDA utilise pure soft PL. None of them consider the potential of a hybrid PL strategy. Hard PL methods also assign their PLs based on direct network prediction, which fail to consider the label relationship information, resulting in error-prone PLs. Besides, due to the ignorance of RS, none of these methods utilise RS to benefit PL assignment. 
  \item From the feature alignment perspective, methods such as JADA, CAFA, CAFD, DCAN and GDCAN minimise probabilistic divergence via Wasserstein distance minimisation, direct divergence loss minimisation, etc. Methods such as ADAR, PDA, JADA, CAFA, CAFD, CDAN etc., focus on domain discrimination-based feature alignment. However, using adversarial learning to facilitate better feature alignment via error knowledge learning, and simultaneously avoid error knowledge forgetfulness is a void that needs to be filled. 
\end{itemize}
Therefore, these state-of-the-art baselines are representative and sufficient to verify the ABRSI method's effectiveness.

{\renewcommand{\arraystretch}{1.2}%
\begin{table}[!t]
  \caption{Intrusion detection performance evaluation using F1-Score of $10$ methods on four randomly selected tasks}
  \vspace{-0.3cm}
  \centering
  \label{tab:f1_score_table}
  \begin{tabular}{c|cccc|c}
  \Xhline{2\arrayrulewidth}
  $\mathcal{D}_{S} \rightarrow \mathcal{D}_{T}$ & C $\rightarrow$ G & N $\rightarrow$ W & K $\rightarrow$ W & K $\rightarrow$ B$^*$ & Avg \\ \hline
  CDAN & 0.49 & 0.53 & 0.56 & 0.37 & 0.49 \\
  MCC & 0.45 & 0.50 & 0.53 & 0.40 & 0.47 \\
  CAFD & 0.60 & 0.56 & 0.51 & 0.39 & 0.50 \\
  JADA & 0.59 & 0.52 & 0.56 & 0.39 & 0.52 \\
  CAFA & 0.61 & 0.54 & 0.54 & 0.39 & 0.52 \\
  DCAN & 0.60 & 0.46 & 0.56 & 0.40 & 0.51 \\
  GDCAN & 0.52 & 0.51 & 0.57 & 0.39 & 0.50 \\
  PDA & 0.67 & 0.59 & 0.57 & 0.34 & 0.54 \\
  ADAR & 0.62 & 0.57 & 0.59 & 0.42 & 0.55 \\ \hline
  \rowcolor{gray}
  \textbf{ABRSI} & \textbf{0.88} & \textbf{0.70} & \textbf{0.70} & \textbf{0.51} & \textbf{0.70} \\ \Xhline{2\arrayrulewidth}
  \end{tabular}
\end{table}}

{\renewcommand{\arraystretch}{1.2}%
\begin{table}[!t]
  \caption{Intrusion detection performance evaluation using AUC score of $9$ methods on four randomly selected tasks}
  \vspace{-0.3cm}
  \centering
  \label{tab:auc_score_table}
  \begin{tabular}{c|cccc|c}
  \Xhline{2\arrayrulewidth}
  $\mathcal{D}_{S} \rightarrow \mathcal{D}_{T}$ & C $\rightarrow$ G & N $\rightarrow$ W & K $\rightarrow$ W & K $\rightarrow$ B$^*$ & Avg \\ \hline
  CDAN & 0.47 & 0.51 & 0.53 & 0.50 & 0.50 \\
  MCC & 0.51 & 0.48 & 0.46 & 0.54 & 0.50 \\
  CAFD & 0.63 & 0.56 & 0.43 & 0.50 & 0.53 \\
  JADA & 0.50 & 0.55 & 0.54 & 0.54 & 0.53 \\
  CAFA & 0.62 & 0.55 & 0.57 & 0.55 & 0.57 \\
  DCAN & 0.51 & 0.44 & 0.55 & 0.58 & 0.52 \\
  GDCAN & 0.50 & 0.49 & 0.58 & 0.56 & 0.53 \\
  PDA & 0.42 & 0.55 & 0.52 & 0.59 & 0.52 \\ \hline
  \rowcolor{gray}
  \textbf{ABRSI} & \textbf{0.86} & \textbf{0.70} & \textbf{0.59} & \textbf{0.65} & \textbf{0.70} \\ \Xhline{2\arrayrulewidth}
  \end{tabular}
  \vspace{-3mm}
\end{table}}

\subsection{Intrusion Detection Performance}\label{sec:intrusion_detection_performance}

We first present the intrusion detection accuracy results on nine randomly selected tasks in Table \ref{tab:giant_performance_table}. As we can observe, the ABRSI model significantly outperforms all state-of-the-art counterparts. Specifically, it outperforms the best comparing method ADAR by $9.2\%$, demonstrating the best intrusion detection effectiveness. It's natural to observe this since the ADAR method fails to use the adaptive bi-recommendation matching mechanism, and ignores the usefulness of \rw{the} RS-based PL assignment strategy. It also lacks sufficient error knowledge learning to promote finer intrusion knowledge transfer, resulting in compromised intrusion detection performance. 

We then present the performance measured using precision and recall in Fig. \ref{fig:figure_metrics_pr}, in which the X-axis and Y-axis indicate precision and recall, respectively. As we can see, the ABRSI method sits at the top-right corner, demonstrating the best precision and recall compared with other methods with a significant performance improvement. The best precision performance indicates the most intrusions flagged by the ABRSI are correct decisions, while the best recall performance means that the ABRSI model can flag as many intrusions as possible. 

We also calculate the F1-score, a harmonic mean that reflects the balance between precision and recall, and present the result in Table \ref{tab:f1_score_table}. The ABRSI method has the best F1-score performance, outperforming the best counterpart PDA and ADAR by $29.6\%$ and $27.3\%$, respectively. The best F1-score indicates that the ABRSI method can balance intrusion flagging and false-alarm avoidance properly, demonstrating its real-world usefulness. 

Finally, we measure the AUC score, an evaluation metric reflecting the intrusion detection capability. The results are presented in Table \ref{tab:auc_score_table}. Note that the ADAR yields ranked recommendation, not a probabilistic prediction, hence, it is not applicable to AUC measurement. The ABRSI shows superior performance compared with all counterparts, yielding a $34.6\%$ AUC improvement, demonstrating the best ability to detect intrusions with an excellent separability. 

Overall speaking, the superior performances of ABRSI on all evaluation metrics indicate that the ABRSI model can detect intrusions as accurate as possible. It flags as many intrusions as possible from all possible malicious behaviours while avoiding triggering too much false alarms. It can balance intrusion flagging and false-alarm avoidance properly, thanks to its best ability of intrusion separability. Therefore, the results demonstrate the real-world applicability of the ABRSI model for effective IoT intrusion detection.

{\renewcommand{\arraystretch}{1.2}%
\begin{table*}[!ht]
  \caption{Ablation Group A: Adaptive Bi-Recommendation Effectiveness}
  \centering
  \label{tab:ablation_group_a_adaptive_bi_recommendation_effectiveness}
  \begin{tabular}{c|cc|cccc|c}
  \Xhline{2\arrayrulewidth}
  \multirow{2}{*}{Experiment} & \multicolumn{2}{c|}{Adaptive Bi-Recommendation Usage} & \multicolumn{4}{c|}{Ablation Results} & \multirow{2}{*}{Avg} \\ \cline{2-7}
   & ABR Matching & RS-based $\text{PL}^{\text{RS}}$ & N $\rightarrow$ W & C $\rightarrow$ G & C $\rightarrow$ B & K $\rightarrow$ B$^*$ &  \\ \hline
  A$_1$ & \xmark & \checkmark\hspace{1mm}($\text{PL}^{\text{NN}}$+$\text{PL}^{\text{RS}}$+$\text{PL}^{\text{SR}}$+$\text{PL}^{\text{TR}}$) & 63.08 & 87.27 & 62.22 & 40.60 & 63.29 \\
  A$_2$ & \checkmark & \xmark\hspace{1mm}($\text{PL}^{\text{NN}}$+$\text{PL}^{\text{SR}}$+$\text{PL}^{\text{TR}}$) & 62.63 & 87.47 & 60.34 & 41.83 & 63.07 \\
  A$_3$ & \xmark & \xmark & 59.56 & 87.13 & 61.56 & 40.23 & 62.12 \\
  \rowcolor{gray}
  \textbf{ABRSI Full} & \checkmark & \checkmark & \textbf{67.45} & \textbf{88.04} & \textbf{65.42} & \textbf{47.42} & \textbf{67.08} \\ \Xhline{2\arrayrulewidth}
  \end{tabular}
\end{table*}}

{\renewcommand{\arraystretch}{1.2}%
\begin{table*}[!ht]
  \caption{Ablation Group B: Pseudo labelling voting component effectiveness}
  \centering
  \label{tab:ablation_group_b_pseudo_labelling_voting_component_effectiveness}
  \begin{tabular}{c|cccc|cccc|c}
  \Xhline{2\arrayrulewidth}
  \multirow{2}{*}{Experiment} & \multicolumn{4}{c|}{PL Voting Components} & \multicolumn{4}{c|}{Ablation Results} & \multirow{2}{*}{Avg} \\ \cline{2-9}
   & $\text{PL}^{\text{NN}}$ & $\text{PL}^{\text{RS}}$ & $\text{PL}^{\text{SR}}$ & $\text{PL}^{\text{TR}}$ & N $\rightarrow$ W & C $\rightarrow$ G & C $\rightarrow$ B & K $\rightarrow$ B$^*$ &  \\ \hline
  B$_1$ & \checkmark & \checkmark & \xmark & \xmark & 59.61 & 86.91 & 63.11 & 41.38 & 62.75 \\
  B$_2$ & \checkmark & \xmark & \checkmark & \xmark & 56.02 & 72.58 & 55.59 & 40.15 & 56.09 \\
  B$_3$ & \checkmark & \xmark & \xmark & \checkmark & 56.10 & 87.06 & 59.32 & 38.08 & 60.14 \\
  \rowcolor{gray}
  \textbf{ABRSI Full} & \checkmark & \checkmark & \checkmark & \checkmark & \textbf{67.45} & \textbf{88.04} & \textbf{65.42} & \textbf{47.42} & \textbf{67.08} \\ \Xhline{2\arrayrulewidth}
  \end{tabular}
  \vspace{-3mm}
\end{table*}}

\subsection{Adaptive Bi-Recommendation Effectiveness}\label{sec:adaptive_bi_recommendation_effectiveness}

We now analyse the effectiveness of the adaptive bi-recommendation mechanism, and the contribution it makes on improving the quality of hard PL assignment. The results are indicated in Table \ref{tab:ablation_group_a_adaptive_bi_recommendation_effectiveness}. The Full ABRSI outperforms all the ABR-ablated counterparts by a large margin. Specifically, the ABR and RS-based hard PL assignment contribute $3.8\%$ and $4\%$ of performance improvement. Besides, completely lacking the adaptive bi-recommendation mechanism degrades the performance by $5\%$. It therefore verifies the positive contribution made by the ABR mechanism, and the usefulness of its constituting components.

\subsection{Self-improving: Pseudo Label Voting Mechanism Effectiveness}\label{sec:pseudo_label_voting_mechanism_effectiveness}

\textbf{Voting Component Effectiveness} We verify the effectiveness of each hard PL voting component, and present the results in Table \ref{tab:ablation_group_b_pseudo_labelling_voting_component_effectiveness}. As we can observe, the full ABRSI shows superior performance than its voting-ablated counterparts by a large margin, demonstrating that all voters are indispensable for an accurate hard PL assignment outcome. 

\begin{figure}[!ht]
  \begin{center}
    \includegraphics[width=0.45\textwidth,keepaspectratio]{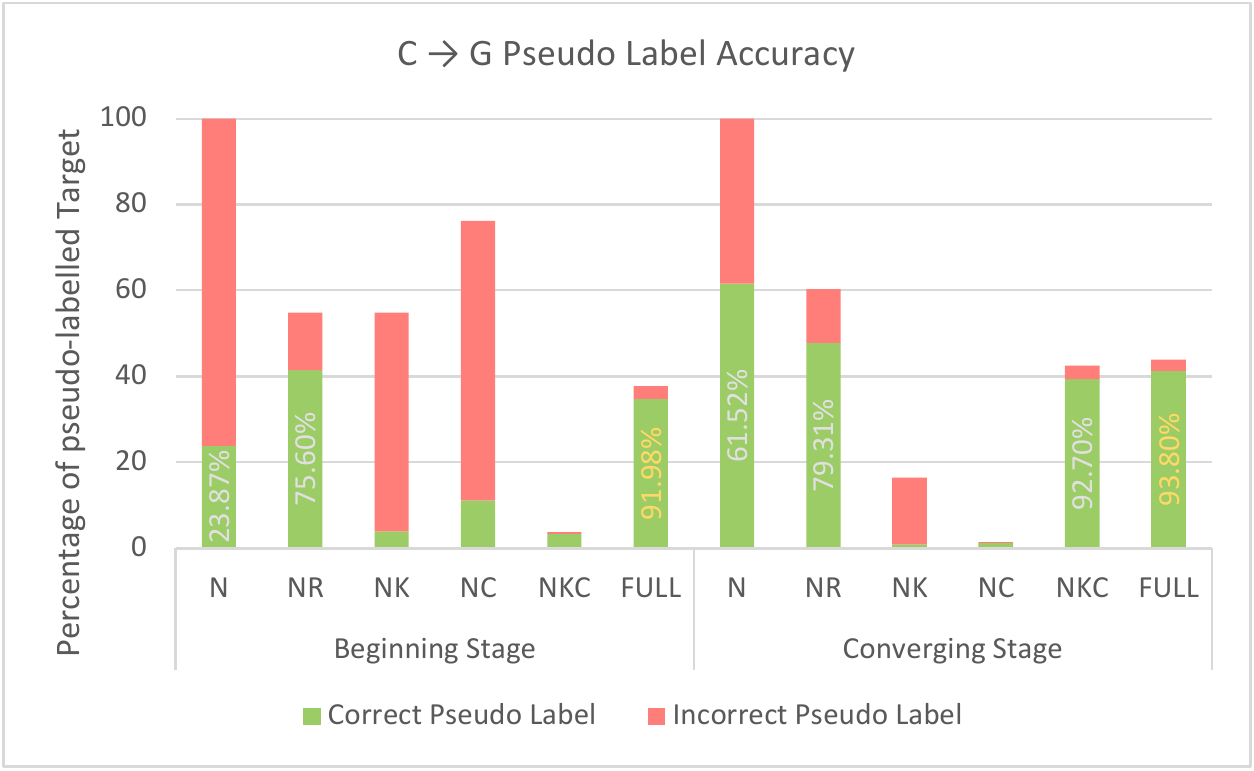}\\
    \caption{Pseudo label accuracy visualisation. The bar height indicates ratio of hard PL assigned target instances, green and red colour indicate correctly and incorrectly assigned PL ratio, respectively}
    \vspace{-0.5cm}
    \label{fig:figure_pseudo_label_accuracy}
  \end{center}
\end{figure}

To further illustrate the hard PL accuracy, we present the hard PL accuracy, hard PL ratio and overall intrusion detection accuracy during the beginning and converging stages in Table \ref{tab:pl_big_table}, and visualise the result of a randomly selected task in Fig. \ref{fig:figure_pseudo_label_accuracy}. As we can observe, the full ABRSI can yield accurate hard PL assignment even at the beginning training stage. It will not blindly increase the percentage of hard PL-assigned instances, since the accuracy of hard PL matters more than the pure amount. As the training progresses, the hard PL accuracy of full ABRSI can even reach $99\%$, the highest among all its ablated counterparts. Besides, the hard PL ratio is also gradually increased. By assigning hard PL in an accurate manner, the full ABRSI leads to the best intrusion detection performance, thanks to the contribution made by each hard PL voting component.

{\renewcommand{\arraystretch}{1.2}%
\begin{table*}[!ht]
  \caption{Hard pseudo label accuracy, ratio and overall prediction accuracy under different pseudo label voting strategies}
  \centering
  \label{tab:pl_big_table}
  \begin{tabular}{c|c|ccc|ccc}
  \Xhline{2\arrayrulewidth}
  \multirow{2}{*}{$\mathcal{D}_{S} \rightarrow \mathcal{D}_{T}$} & \multirow{2}{*}{PL Strategy} & \multicolumn{3}{c|}{Beginning Stage} & \multicolumn{3}{c}{Converging Stage} \\ \cline{3-8} 
   &  & Hard PL Acc & Hard PL Ratio & Overall Acc & Hard PL Acc & Hard PL Ratio & Overall Acc \\ \hline
  \multirow{7}{*}{N $\rightarrow$ W} & $\text{PL}^{\text{NN}}$ & 46.42 & 100.00 & 46.42 & 51.00 & 100.00 & 51.00 \\
   & $\text{PL}^{\text{NN}}$ Soft & 0.00 & 0.00 & 51.10 & 0.00 & 0.00 & 54.45 \\
   & $\text{PL}^{\text{NN}}$+$\text{PL}^{\text{RS}}$ & 79.31 & 3.62 & 52.12 & 58.40 & 62.08 & 61.21 \\
   & $\text{PL}^{\text{NN}}$+$\text{PL}^{\text{SR}}$ & 22.93 & 15.86 & 50.10 & 53.71 & 67.64 & 55.03 \\
   & $\text{PL}^{\text{NN}}$+$\text{PL}^{\text{TR}}$ & 47.02 & 89.56 & 46.79 & 55.09 & 27.02 & 56.18 \\
   & $\text{PL}^{\text{NN}}$+$\text{PL}^{\text{SR}}$+$\text{PL}^{\text{TR}}$ & 51.82 & 64.61 & 54.65 & 63.84 & 45.27 & 63.73 \\
   & \cellcolor{gray}\textbf{FULL} & \cellcolor{gray}76.62 & \cellcolor{gray}8.13 & \cellcolor{gray}52.76 & \cellcolor{gray}\textbf{98.92} & \cellcolor{gray}14.44  & \cellcolor{gray}\textbf{71.02} \\ \hline
  \multirow{7}{*}{C $\rightarrow$ G} & $\text{PL}^{\text{NN}}$ & 23.87 & 100.00 & 23.87 & 61.52 & 100.00 & 61.52 \\
   & $\text{PL}^{\text{NN}}$ Soft & 0.00 & 0.00 & 41.24 & 0.00 & 0.00 & 80.42 \\
   & $\text{PL}^{\text{NN}}$+$\text{PL}^{\text{RS}}$ & 75.60 & 54.87 & 78.10 & 79.31 & 60.36 & 85.38 \\
   & $\text{PL}^{\text{NN}}$+$\text{PL}^{\text{SR}}$ & 9.60 & 40.64 & 53.90 & 5.80 & 16.39 & 72.49 \\
   & $\text{PL}^{\text{NN}}$+$\text{PL}^{\text{TR}}$ & 14.68 & 76.17 & 32.83 & 91.07 & 1.40 & 86.48 \\
   & $\text{PL}^{\text{NN}}$+$\text{PL}^{\text{SR}}$+$\text{PL}^{\text{TR}}$ & 90.55 & 3.85 & 53.20 & 92.70 & 42.54 & 87.18 \\
   & \cellcolor{gray}\textbf{FULL} & \cellcolor{gray}91.98 & \cellcolor{gray}37.76 & \cellcolor{gray}77.90 & \cellcolor{gray}\textbf{93.80} & \cellcolor{gray}43.99 & \cellcolor{gray}\textbf{88.54} \\ \Xhline{2\arrayrulewidth}
  \end{tabular}
  \vspace{-3mm}
\end{table*}}

{\renewcommand{\arraystretch}{1.2}%
\begin{table}[!ht]
  \caption{Hard PL diversity measured using Hellinger Distance}
  \centering
  \label{tab:hard_pl_diversity}
  \begin{tabular}{c|cccc|c}
  \Xhline{2\arrayrulewidth}
  Experiment & N $\rightarrow$ W & C $\rightarrow$ G & C $\rightarrow$ B & K $\rightarrow$ B$^*$ & Avg \\ \hline
  Hard $\text{PL}^{\text{NN}}$ & 0.26 & 0.20 & 0.38 & 0.57 & 0.35 \\
  \rowcolor{gray}
  \textbf{ABRSI Full} & \textbf{0.20} & \textbf{0.15} & \textbf{0.08} & \textbf{0.47} & \textbf{0.23} \\ \Xhline{2\arrayrulewidth}
  \end{tabular}
\end{table}}

\textbf{Hard PL Diversity} We also find out that the ABRSI can yield hard PL assignment with a higher diversity than the hard-PL-only assignment without voting. We use Hellinger Distance to measure the diversity, i.e., the distance between the hard PL distribution and the ideal distribution of target II domain. The smaller the Hellinger Distance is, the better the diversity. The Hellinger Distance has the following definition: 
\begin{equation}
  H(P_{PL}, P_{T}) = \frac{1}{\sqrt{2}} \sqrt{\sum_{k = 1}^{K} (\sqrt{P_{PL}^{(k)}} - \sqrt{P_{T}^{(k)}})^2}\,,
\end{equation}
where $P_{PL}$ and $P_{T}$ denote the hard PL distribution and target II intrusion distribution, $P_{PL}^{(k)}$ denotes the $k$\textsuperscript{th} element of $P_{PL}$. The results are shown in Table \ref{tab:hard_pl_diversity}. As we can observe, the full ABRSI yields an improved hard PL diversity \rw{compared with} its hard-PL-only-based counterpart by reducing the Hellinger Distance by $34.3\%$. The reason is that the hard-PL-only-based counterpart is more likely to focus on easy-to-predict categories without the refinement achieved by the voting mechanism. Conversely, the full ABRSI is assisted by the hard PL voting mechanism, which can improve the hard PL accuracy and meanwhile reduce the over-emphasis on easy-to-predict categories, which leads to better hard PL diversity, and hence benefits the intrusion detection effectiveness.

\subsection{Self-improving: Hybrid Pseudo Label Effectiveness}\label{sec:hybrid_pseudo_label_effectiveness}

We now verify the superiority \rw{of the} ABRSI model compared with its hard-PL-only and soft-PL-only counterpart and present the results in Table \ref{tab:ablation_group_c_hybrid_pseudo_labelling_effectiveness}. As we can notice, the full ABRSI model has the best intrusion detection accuracy, benefitted by the hybrid PL strategy. Its hard-PL-only and soft-PL-only peer show a $16.9\%$ and $9.7\%$ accuracy drop, demonstrating that both components constituting the hybrid PL strategy are indispensable.

{\renewcommand{\arraystretch}{1.2}%
\begin{table*}[!t]
  \caption{Ablation Group C: Hybrid pseudo labelling effectiveness}
  \centering
  \label{tab:ablation_group_c_hybrid_pseudo_labelling_effectiveness}
  \begin{tabular}{c|cc|cccc|c}
  \Xhline{2\arrayrulewidth}
  \multirow{2}{*}{Experiment} & \multicolumn{2}{c|}{PL Types} & \multicolumn{4}{c|}{Ablation Results} & \multirow{2}{*}{Avg} \\ \cline{2-7}
   & Involving Hard PL & Involving Soft PL & N $\rightarrow$ W & C $\rightarrow$ G & C $\rightarrow$ B & K $\rightarrow$ B$^*$ &  \\ \hline
  C$_1$ & \checkmark & \xmark & 50.92 & 61.52 & 52.39 & 35.86 & 50.17 \\
  C$_2$ & \xmark & \checkmark & 53.74 & 80.42 & 55.37 & 40.05 & 57.40 \\
  \rowcolor{gray}
  \textbf{ABRSI Full} & \checkmark & \checkmark & \textbf{67.45} & \textbf{88.04} & \textbf{65.42} & \textbf{47.42} & \textbf{67.08} \\ \Xhline{2\arrayrulewidth}
  \end{tabular}
\end{table*}}

\subsection{Self-improving: Soft PL Diversity and Confidence}\label{sec:soft_pl_diversity_and_confidence}

We now verify the effectiveness of soft PL diversity and confidence improvement, part of the self-improving mechanism. The intrusion detection accuracy results are presented in Table \ref{tab:ablation_group_d_soft_pseudo_label_confidence_and_diversity}. The diversity and confidence improvement contribute $11.6\%$ and $4.1\%$ of accuracy improvement, respectively. Without both improvements, the intrusion detection accuracy will even drop by $12\%$. 

Besides, the Hellinger Distance is also used to measure the diversity of soft PL with results presented in Table \ref{tab:soft_pl_diversity}, and the confidence is measured using \rw{the} percentage of high certainty predictions with certainty threshold set to $0.7$ and the results are given in Table \ref{tab:soft_pl_certainty}. As we can observe, the diversity improvement component decreases \rw{the} Hellinger Distance by $50\%$, and the confidence improvement component achieves higher prediction certainty, with $36\%$ of the intrusion detection decision having a certainty greater than $70\%$. Hence, these self-improving components achieve a globally diverse and individually certain soft PL assignment, which boosts the intrusion detection performance. 

{\renewcommand{\arraystretch}{1.2}%
\begin{table*}[!ht]
  \caption{Ablation Group D: Soft pseudo label confidence and diversity self-improving components}
  \centering
  \label{tab:ablation_group_d_soft_pseudo_label_confidence_and_diversity}
  \begin{tabular}{c|cc|cccc|c}
  \Xhline{2\arrayrulewidth}
  \multirow{2}{*}{Experiment} & \multicolumn{2}{c|}{Soft PL confidence and diversity improvement component} & \multicolumn{4}{c|}{Ablation Results} & \multirow{2}{*}{Avg} \\ \cline{2-7}
   & Confidence Improvement ($\mathcal{L}_{TE}$) & Diversity Improvement ($\mathcal{L}_{DIV}$) & N $\rightarrow$ W & C $\rightarrow$ G & C $\rightarrow$ B & K $\rightarrow$ B$^*$ &  \\ \hline
  D$_1$ & \xmark & \checkmark & 58.88 & 87.80 & 64.00 & 41.21 & 62.97 \\
  D$_2$ & \checkmark & \xmark & 55.77 & 69.47 & 57.47 & 39.06 & 55.44 \\
  D$_3$ & \xmark & \xmark & 54.95 & 76.83 & 53.09 & 35.64 & 55.13 \\
  \rowcolor{gray}
  \textbf{ABRSI Full} & \checkmark & \checkmark & \textbf{67.45} & \textbf{88.04} & \textbf{65.42} & \textbf{47.42} & \textbf{67.08} \\ \Xhline{2\arrayrulewidth}
  \end{tabular}
  \vspace{-3mm}
\end{table*}}

{\renewcommand{\arraystretch}{1.2}%
\begin{table}[!ht]
  \caption{Soft pseudo label diversity measured using Hellinger Distance}
  \centering
  \label{tab:soft_pl_diversity}
  \begin{tabular}{c|cccc|c}
  \Xhline{2\arrayrulewidth}
  Experiment & N $\rightarrow$ W & C $\rightarrow$ G & C $\rightarrow$ B & K $\rightarrow$ B$^*$ & Avg \\ \hline
  ABRSI$-\mathcal{L}_{TE}$ & 0.27 & 0.32 & 0.41 & 0.43 & 0.36 \\
  \rowcolor{gray}
  \textbf{ABRSI Full} & \textbf{0.13} & \textbf{0.04} & \textbf{0.19} & \textbf{0.36} & \textbf{0.18} \\ \Xhline{2\arrayrulewidth}
  \end{tabular}
\end{table}}

{\renewcommand{\arraystretch}{1.2}%
\begin{table}[!ht]
  \caption{Soft pseudo label certainty measured using percentage of high certainty predictions (probability threshold$=70\%$)}
  \centering
  \label{tab:soft_pl_certainty}
  \begin{tabular}{c|cccc|c}
  \Xhline{2\arrayrulewidth}
  Experiment & N $\rightarrow$ W & C $\rightarrow$ G & C $\rightarrow$ B & K $\rightarrow$ B$^*$ & Avg \\ \hline
  ABRSI$-\mathcal{L}_{DIV}$ & 0.21 & 0.00 & 0.81 & 0.01 & 0.26 \\
  \rowcolor{gray}
  \textbf{ABRSI Full} & \textbf{0.25} & \textbf{0.05} & \textbf{0.97} & \textbf{0.06} & \textbf{0.33} \\ \Xhline{2\arrayrulewidth}
  \end{tabular}
\end{table}}

\subsection{Error Knowledge Learning Necessity and Effectiveness}\label{sec:error_knowledge_learning_usefulness_and_effectiveness}

\textbf{Error Knowledge Learning Necessity} We verify the necessity of the EKL by comparing it with three counterparts: $E_1$, which ablates the EKL; $E_2$, which uses a domain discriminator that distinguishes the domain origin of instances; $E_3$, which directly minimises the Euclidean distance between probabilistic outputs. The results are illustrated in Table \ref{tab:ablation_group_e_error_knowledge_learning_necessity}. As we can see, the ABRSI achieves the best intrusion detection accuracy with a $6.7\%$ accuracy improvement on average. It is natural to observe this since both $E_1$ and $E_2$ lacks consideration of error knowledge mining and elimination, while $E_3$ eliminates error knowledge in a coarse-grained manner. It in turn verifies the necessity of the EKL in the ABRSI model.

{\renewcommand{\arraystretch}{1.2}%
\begin{table*}[b]
  \caption{Ablation Group E: Error knowledge learning necessity analysis}
  \centering
  \label{tab:ablation_group_e_error_knowledge_learning_necessity}
  \begin{tabular}{c|c|c|cccc|c}
  \Xhline{2\arrayrulewidth}
  \multirow{2}{*}{Experiment} & \multirow{2}{*}{Involve Discriminator} & \multirow{2}{*}{Description} & \multicolumn{4}{c|}{Ablation Results} & \multirow{2}{*}{Avg} \\ \cline{4-7}
   &  &  & N $\rightarrow$ W & C $\rightarrow$ G & C $\rightarrow$ B & K $\rightarrow$ B$^*$ &  \\ \hline
  E$_1$: $\gamma = 0$ & \checkmark & Without EKL & 64.21 & 87.76 & 62.18 & 40.52 & 63.67 \\
  E$_2$: Domain Dis & \checkmark & With domain discriminator & 60.47 & 87.50 & 62.20 & 41.14 & 62.83 \\
  E$_3$: Prob Matching & \xmark & With probabilistic output matching & 54.74 & 65.32 & 58.35 & 39.65 & 54.52 \\
  \rowcolor{gray}
  \textbf{ABRSI Full} & \checkmark & \textbf{With EKL} & \textbf{67.45} & \textbf{88.04} & \textbf{65.42} & \textbf{47.42} & \textbf{67.08} \\ \Xhline{2\arrayrulewidth}
  \end{tabular}
\end{table*}}

{\renewcommand{\arraystretch}{1.2}%
\begin{table*}[t]
  \caption{Ablation Group F: Error knowledge learning constituting component effectiveness analysis}
  \centering
  \label{tab:ablation_group_f_error_knowledge_learning_constituting_component_effectiveness}
  \begin{tabular}{c|cc|cccc|c}
  \Xhline{2\arrayrulewidth}
  \multirow{2}{*}{Experiment} & \multicolumn{2}{c|}{EKL} & \multicolumn{4}{c|}{Ablation Results} & \multirow{2}{*}{Avg} \\ \cline{2-7}
   & $EK_{\mathcal{P}}$ & $EK_{\mathcal{R}}$ & N$\rightarrow$W & C$\rightarrow$G & C$\rightarrow$B & K$\rightarrow$B$^*$ & \\ \hline
  F$_1$ & \xmark & \checkmark & 59.31 & 87.45 & 62.84 & 40.77 & 62.59 \\
  F$_2$ & \checkmark & \xmark & 60.72 & 87.43 & 62.34 & 42.00 & 63.12 \\
  \rowcolor{gray}
  \textbf{ABRSI Full} & \checkmark & \checkmark & \textbf{67.45} & \textbf{88.04} & \textbf{65.42} & \textbf{47.42} & \textbf{67.08} \\ \Xhline{2\arrayrulewidth}
  \end{tabular}
\end{table*}}

\begin{figure*}[!t]
  \begin{center}
    \includegraphics[width=\textwidth,keepaspectratio]{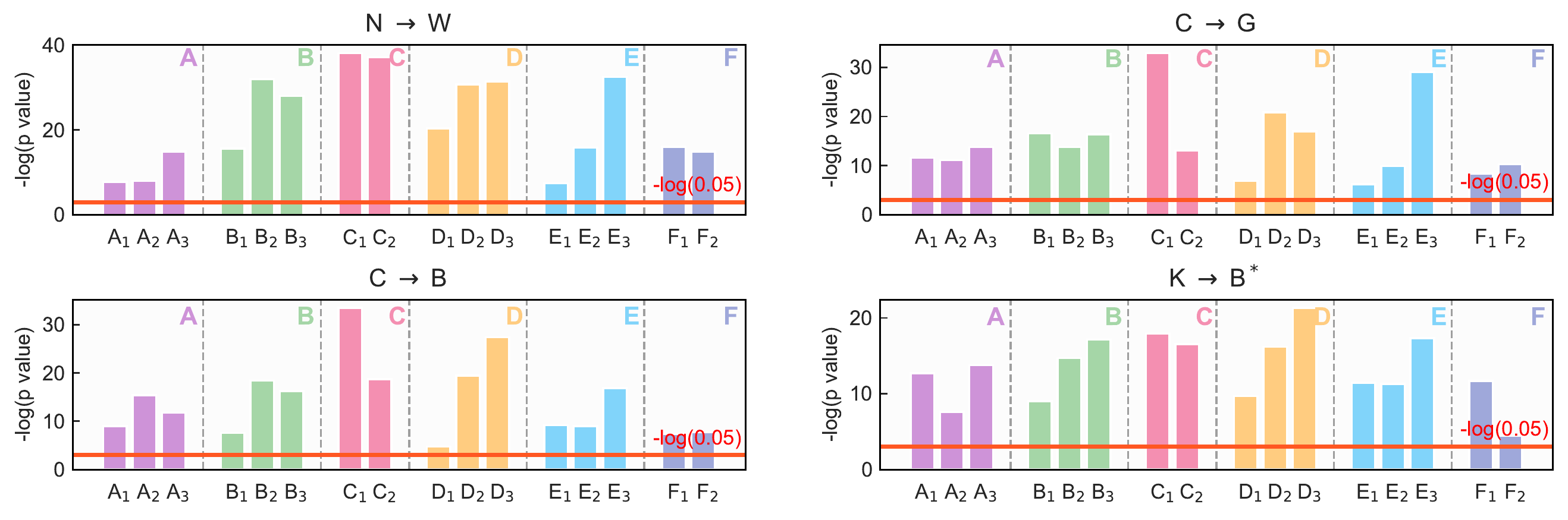}\\
    \vspace{-3mm}
    \caption{Hypothesis testing for ablation group A to F. The red line denotes the significance threshold $-log(0.05)$}
    \vspace{-0.5cm}
    \label{fig:figure_hypothesis_testing}
  \end{center}
\end{figure*}

\begin{figure*}[t]
  \begin{center}
    \includegraphics[width=0.77\textwidth,keepaspectratio]{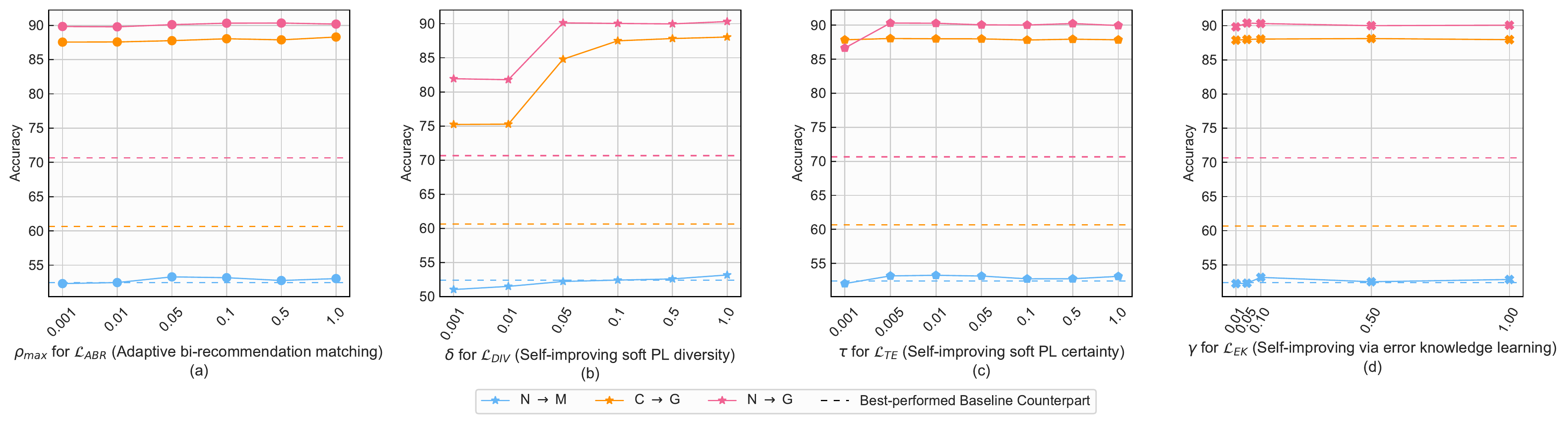}\\
    \vspace{-3mm}
    \caption{Parameter sensitivity verification for balancing hyperparameter $\rho$, $\delta$, $\tau$ and $\gamma$ in the overall optimisation objective}
    \vspace{-0.5cm}
    \label{fig:figure_parameter_sensitivity_1}
  \end{center}
\end{figure*}

\begin{figure*}[t]
  \begin{center}
    \includegraphics[width=\textwidth,keepaspectratio]{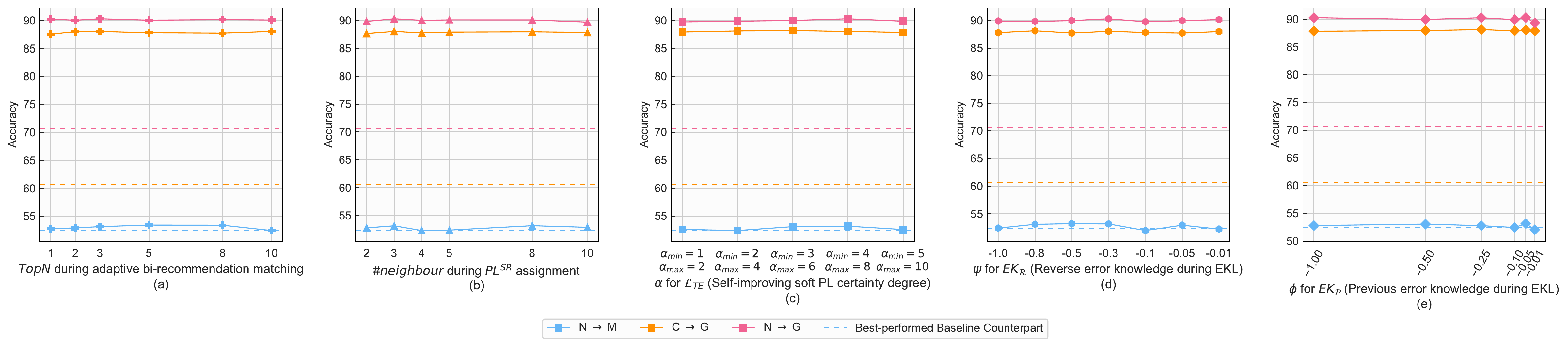}\\
    \vspace{-3mm}
    \caption{Parameter sensitivity verification for other hyperparameters}
    \vspace{-0.7cm}
    \label{fig:figure_parameter_sensitivity_2}
  \end{center}
\end{figure*}

\textbf{EKL Constituting Component Effectiveness} We verify the necessity of error knowledge learning assisted by previous error knowledge $EK_{\mathcal{P}}$ and reverse error knowledge $EK_{\mathcal{R}}$. As shown in Table \ref{tab:ablation_group_f_error_knowledge_learning_constituting_component_effectiveness}, the $EK_{\mathcal{P}}$ and $EK_{\mathcal{R}}$ contribute $4.5\%$ and $4\%$ intrusion accuracy improvement. Hence, it verifies that the reverse error knowledge can effectively strengthen the error knowledge elimination, while the error knowledge forgetfulness is effectively prevented by the previous error knowledge learning, leading to more accurate intrusion detection.

\subsection{Hypothesis Testing of Contributions}\label{sec:hypothesis_testing_of_contributions}

We perform significance T-test \rw{with the} significance threshold \rw{of} $0.05$ to verify the statistical significance of the contributions made by each ABRSI constituting component. The results are illustrated in Fig. \ref{fig:figure_hypothesis_testing}. The red horizontal line indicates the significance threshold $-log(0.05)$, the higher the bar is beyond the red horizontal line, the more significant is the contribution made by that ablated component. We can observe that all bars are significantly higher than the red line, demonstrating statistical soundness of the contributions made by each constituting component of the ABRSI.

\subsection{Hyperparameter Sensitivity Analysis}\label{sec:parameter_sensitivity_analysis}

We also verify the stability and robustness of the ABRSI model towards hyperparameter settings. The parameter sensitivity results of loss balancing hyperparameters are illustrated in Fig. \ref{fig:figure_parameter_sensitivity_1}, and the parameter sensitivity of all other hyperparameters are presented in Fig. \ref{fig:figure_parameter_sensitivity_2}. The coloured dash lines indicate the corresponding best-performed baseline counterpart for each task. As we can observe, the ABRSI presents very stable performance when hyperparameters vary within their reasonable ranges. Besides, the ABRSI stably outperforms its best baseline counterpart, hence, it verifies the ABRSI performs stably under varied hyperparameter settings. 

Besides, when tackling different data domains that present heterogeneities, all experiments share a fixed set of hyperparameter setting, and the ABRSI model is still capable to achieve satisfying results without the need to constantly reset hyperparameters. Hence, it further demonstrates the robustness of ABRSI in terms of hyperparameter setting.

\subsection{Intrusion Detection Efficiency}\label{sec:intrusion_detection_efficiency}

Finally, we evaluate the computational efficiency of the ABRSI model from two aspects, i.e., the training time taken by each training epoch, and the inference time used when performing intrusion detection on each instance. The results are presented in Table \ref{tab:training_time_per_epoch} and \ref{tab:inference_time_per_instance}, respectively. Note that only the top-2 best-performed baselines are compared for computational efficiency. As we can observe, although the ABRSI model is not the fastest in terms of training speed, its training speed is not significantly slow. Besides, the training can be done on devices with strong computational capability. Therefore, the training efficiency of ABRSI is acceptable. On the other hand, the ABRSI achieves the fastest intrusion detection speed, which demonstrates its usefulness when performing intrusion detection for IoT scenarios. Overall, the ABRSI only involves a neural network structure that is relatively shallow, the recommender system it utilised is not neural network-based and has a low computational complexity compared with other NN-based RS counterparts. \rw{Therefore, the satisfying computational efficiency of the ABRSI model makes its deployment to devices such as network gateways feasible, and hence is practical for real-world IoT intrusion detection. }

\section{Conclusion}\label{sec:conclusion}

In this paper, we form the IoT intrusion detection problem as an unsupervised heterogeneous domain adaptation problem that transfers enriched intrusion knowledge from a network intrusion domain to facilitate better IoT intrusion detection under data-scarcity. Firstly, an adaptive bi-recommendation matching mechanism matches bi-recommendation interests and in turn enforces intrusion feature alignment. They can mutually benefit each other, forming a positive loop that can facilitate finer intrusion knowledge transfer. A self-improving mechanism with four components is also utilised. Specifically, a hard pseudo label voting mechanism aims to improve the accuracy and diversity of hard PL assignment. A hybrid PL strategy introduces better participation and diversity, while preserving enough emphasis on confident and correct predictions. The diversity and certainty of soft PL is also jointly improved to achieve globally diverse and individually certain soft PL. Finally, an error knowledge learning mechanism is used to exploit factors that cause prediction ambiguities, and adaptively conduct error knowledge exploitation, elimination and error knowledge forgetfulness prevention. Holistically, these components form the ABRSI model that can boost intrusion detection performance under the IoT scenarios. Comprehensive experiments on five widely-adopted intrusion detection datasets are conducted and compared against nine state-of-the-art baselines to verify the effectiveness of the ABRSI model. The contribution of each ABRSI constituting component is also verified with statistical significance. Finally, the computational efficiency is examined to show the ABRSI's applicability.

{\renewcommand{\arraystretch}{1.2}%
\begin{table*}[!ht]
  \caption{Computational efficiency: Training time per epoch (second)}
  \vspace{-0.3cm}
  \centering
  \label{tab:training_time_per_epoch}
  \begin{tabular}{c|cccccc|c}
  \Xhline{2\arrayrulewidth}
  $\mathcal{D}_{S} \rightarrow \mathcal{D}_{T}$ & N $\rightarrow$ B$^*$ & N $\rightarrow$ G & N $\rightarrow$ M & C $\rightarrow$ G & K $\rightarrow$ B$^*$ & K $\rightarrow$ W & Avg \\ \hline
  PDA & 1.67 & 1.73 & 1.67 & 1.72 & 1.74 & 1.67 & 1.7 \\
  ADAR & 0.17 & 0.10 & 0.07 & 0.07 & 0.17 & 0.09 & 0.11 \\ \hline
  \rowcolor{gray}
  ABRSI & 0.36 & 0.19 & 0.21 & 0.20 & 0.35 & 0.21 & 0.25 \\ \Xhline{2\arrayrulewidth}
  \end{tabular}
\end{table*}}

{\renewcommand{\arraystretch}{1.2}%
\begin{table*}[!ht]
  \caption{Computational efficiency: Inference time per instance (millisecond = $10^{-3}$ second)}
  \vspace{-0.3cm}
  \centering
  \label{tab:inference_time_per_instance}
  \begin{tabular}{c|cccccc|c}
  \Xhline{2\arrayrulewidth}
  $\mathcal{D}_{S} \rightarrow \mathcal{D}_{T}$ & N $\rightarrow$ B$^*$ & N $\rightarrow$ G & N $\rightarrow$ M & C $\rightarrow$ G & K $\rightarrow$ B$^*$ & K $\rightarrow$ W & Avg \\ \hline
  PDA & 1.12 & 1.09 & 1.25 & 1.09 & 1.13 & 1.06 & 1.12 \\
  ADAR & $2.69 \times 10^{-2}$ & $1.75 \times 10^{-2}$ & $1.68 \times 10^{-2}$ & $1.72 \times 10^{-2}$ & $2.61 \times 10^{-2}$ & $1.72 \times 10^{-2}$ & $2.03 \times 10^{-2}$ \\ \hline
  \rowcolor{gray}
  \textbf{ABRSI} & $\mathbf{2.56 \times 10^{-4}}$ & $\mathbf{1.77 \times 10^{-4}}$ & $\mathbf{2.68 \times 10^{-4}}$ & $\mathbf{1.83 \times 10^{-4}}$ & $\mathbf{2.08 \times 10^{-4}}$ & $\mathbf{2.30 \times 10^{-4}}$ & $\mathbf{2.20 \times 10^{-4}}$ \\ \Xhline{2\arrayrulewidth}
  \end{tabular}
  \vspace{-3mm}
\end{table*}}


\ifCLASSOPTIONcompsoc
  \section*{Acknowledgments}
\else
  \section*{Acknowledgment}
\fi

This work is supported by the Third Xinjiang Scientific Expedition Program (Grant No.2021xjkk1300), and also in part by Science and Technology Development Fund of Macao SAR (FDCT) (1058No.0015/2019/AKP) and the Chinese Academy of Sciences President’s International Fellowship Initiative (Grant No. 2023VTA0001).

\ifCLASSOPTIONcaptionsoff
  \newpage
\fi


\bibliographystyle{IEEEtran}
\bibliography{ABRSI}



\vspace{-10mm}
\begin{IEEEbiography}[{\includegraphics[width=0.9in,height=1.13in,clip,keepaspectratio]{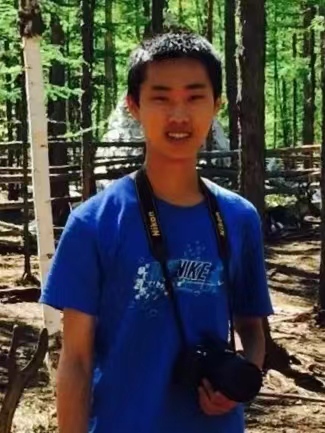}}]{Jiashu Wu} received BSc. degree in Computer Science and Financial Mathematics \& Statistics from the University of Sydney, Australia (2018), and M.IT. degree in Artificial Intelligence from the University of Melbourne, Australia (2020). He is currently pursuing his Ph.D. at the University of Chinese Academy of Sciences (Shenzhen Institute of Advanced Technology, Chinese Academy of Sciences). His research interests including machine learning and cloud computing. 
\end{IEEEbiography}
\vspace{-10mm}
\begin{IEEEbiography}[{\includegraphics[width=0.9in,height=1.13in,clip,keepaspectratio]{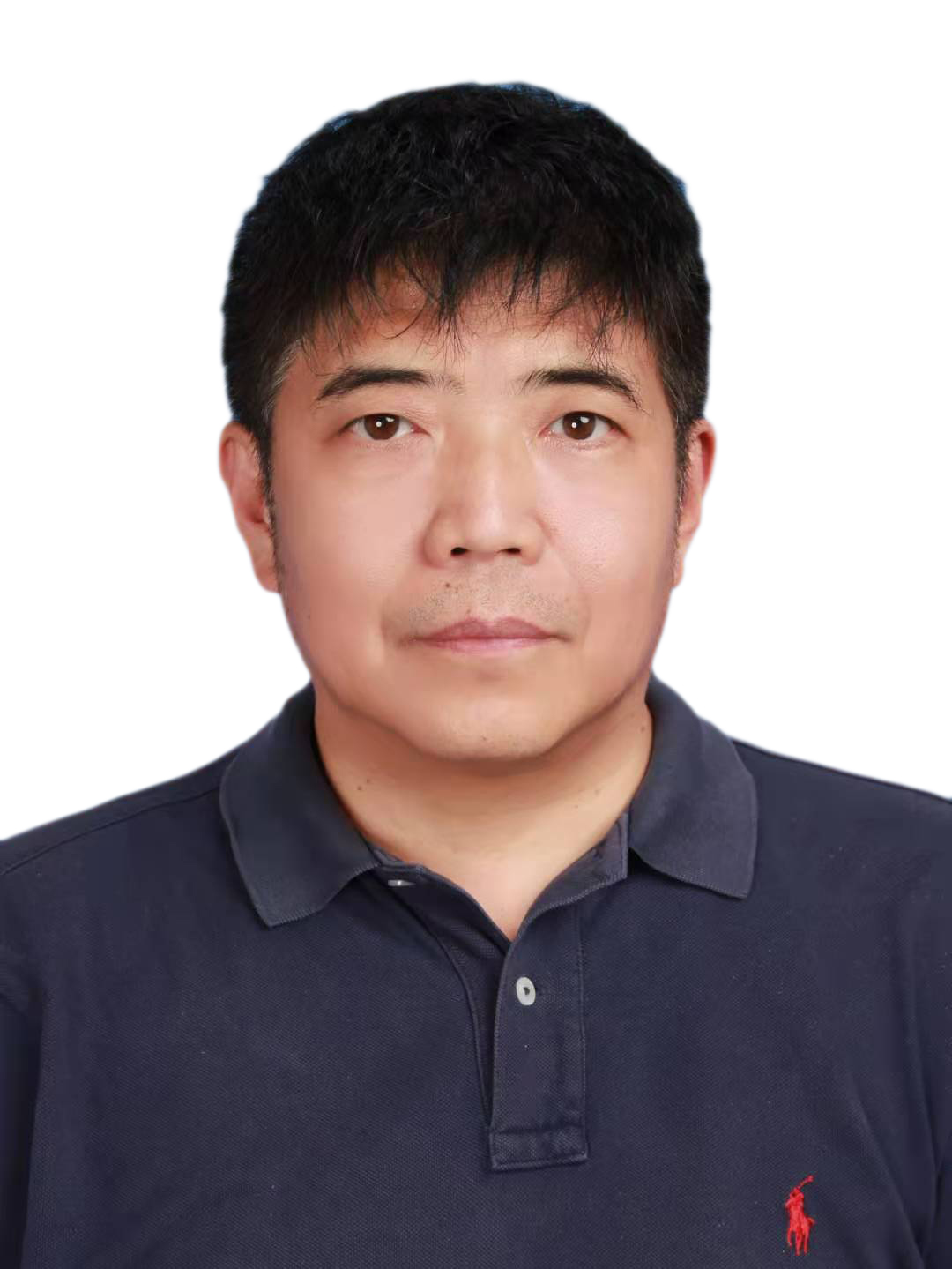}}]{Yang Wang} received the BSc. degree in applied mathematics from Ocean University of China (1989), and the MSc. and Ph.D. degrees in computer science from Carlton University (2001) and University of Alberta, Canada (2008), respectively. He is currently with Shenzhen Institutes of Advanced Technology, Chinese Academy of Sciences, as a professor and with Xiamen University as an adjunct professor. His research interests include service and cloud computing, programming language implementation, and software engineering. He is an Alberta Industry R\&D Associate (2009-2011), and a Canadian Fulbright Scholar (2014-2015). 
\end{IEEEbiography}
\vspace{-10mm}
\begin{IEEEbiography}[{\includegraphics[width=1in,height=1.1in,clip,keepaspectratio]{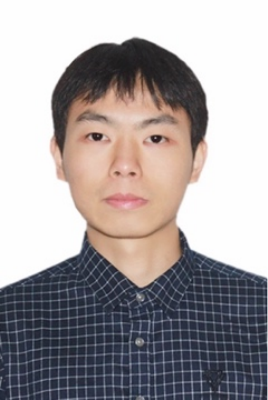}}]{Hao Dai} received the BSc. and M.Sc degrees in Communication and Electronic Technology from the Wuhan University of Technology in 2015 and 2017, respectively. He is currently working toward the Ph.D. degree in the Shenzhen Institute of Advanced Technology, Chinese Academy of Sciences. His research interests include mobile edge computing, federated learning and deep reinforcement learning. 
\end{IEEEbiography}
\vspace{-10mm}
\begin{IEEEbiography}[{\includegraphics[width=1in,height=1.1in,clip,keepaspectratio]{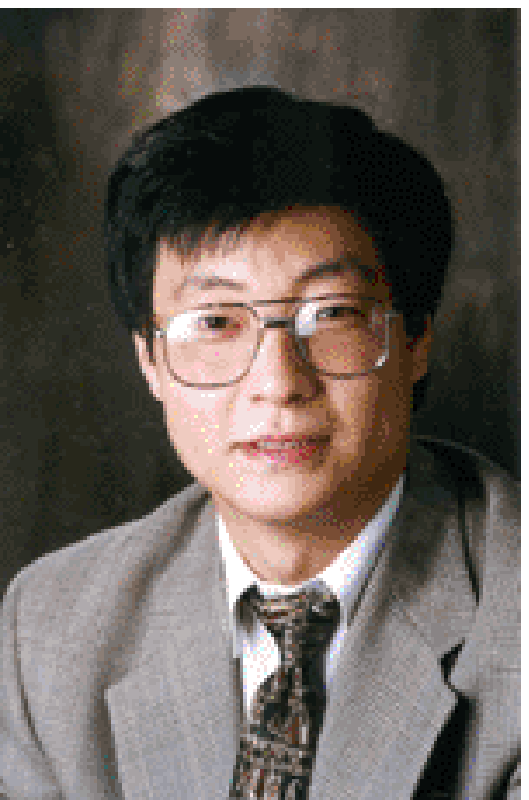}}]{Chengzhong Xu} received the Ph.D. degree from the University of Hong Kong in 1993. He is currently the Dean of Faculty of Science and Technology, University of Macau, China, and the Director of the Institute of Advanced Computing and Data Engineering, Shenzhen Institutes of Advanced Technology of Chinese Academy of Sciences.His research interest includes parallel and distributed systems, service and cloud computing, and software engineering. He has published more than 200 papers in journals and conferences. He serves on a number of journal editorial boards, including IEEE TC, IEEE TPDS, IEEE TCC, JPDC and China Science Information Sciences. He is a fellow of the IEEE.
\end{IEEEbiography}
\vspace{-10mm}
\begin{IEEEbiography}[{\includegraphics[width=0.9in,height=1.2in,clip,keepaspectratio]{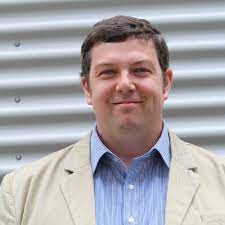}}]{Kenneth B. Kent} has been a Professor in the Faculty of Computer Science since 2002. He is also the Director of the Center for Advanced Studies – Atlantic. Dr. Kent has supervised over 70 graduate students, postdocs, and researchers and has published more than 150 refereed journal articles, conference papers and patents. Dr. Kent is an Honorary Professor at Hochschule Bonn-Rhein-Sieg, where he is also involved in research through the Institute for Visual Computing and the Department of Computer Science. 
\end{IEEEbiography}
\vspace{-10mm}

\clearpage

\section*{Appendix}

\textbf{Acronym Table and Notation Table}\label{sec:acronym_table_and_notation_table}

We provide an acronym table and a notation table for better readability. 

{\renewcommand{\arraystretch}{1.2}%
\begin{table}[!ht]
    \vspace{-5mm}
    \caption{The acronym table}
    \centering
    \label{tab:acronym_table}
    \begin{tabular}{c|l}
    \Xhline{2\arrayrulewidth}
    Acronym & Interpretation \\ \hline
    ABRSI & Adaptive Bi-Recommendation and Self-Improving Network \\
    HDA & Heterogeneous Domain Adaptation \\
    IID & IoT Intrusion Detection \\
    NID & Network Intrusion Detection \\
    DA & Domain Adaptation \\
    NI & Network Intrusion \\
    II & IoT Intrusion \\
    RS(s) & Recommender System(s) \\
    PL(s) & Pseudo Label(s) \\
    ABR & Adaptive Bi-Recommendation Matching Mechanism \\
    SI & Self-Improving Mechanism \\
    EK & Error Knowledge \\
    EKL & Error Knowledge Learning \\
    NN & Neural Network \\
    SR & Source label Relationship \\
    TR & Target label Relationship \\
    \Xhline{2\arrayrulewidth}
    \end{tabular}
    \vspace{-3mm}
\end{table}}

\makeatletter
\setlength{\@fptop}{0pt}
\makeatother

{\renewcommand{\arraystretch}{1.2}%
\begin{table}[!ht]
    \caption{The notation table}
    \centering
    \label{tab:notation_table}
    \begin{tabular}{l|p{6.5cm}}
    \Xhline{2\arrayrulewidth}
    Notation & Interpretation \\ \hline
    $\mathcal{D}_{*}$ & Domain, $* \in \{S, T\}$ \\
    $x_{S_i}$ & $i$\textsuperscript{th} instance in source domain \\
    $y_{S_i}$ & The intrusion category label of the $i$\textsuperscript{th} source instance \\
    $n_{S}$ & Number of source domain instances \\
    $d_{S}$ & Dimension of source domain \\
    $E_{S}(x_i)$ & The source projector \\
    $f(x_i)$ & Features projected by the projector \\
    $d_{C}$ & Dimension of common feature space \\
    $RS_{S}$ & Source-trained recommender system \\
    $M_{S}$ & Original feature matrix of source domain \\
    $R$ & Reduced dimension parameter during LSI \\
    $U^R$ & The feature-latent matrix with reduced dimension $R$ \\
    $T^R$ & The latent transfer matrix with reduced dimension $R$ \\
    $V^{R\top}$ & The instance-latent matrix with reduced dimension $R$ \\
    $x_S^{i'}$ & Transformed source instance $x_S^i$ by the LSI algorithm \\
    $RS_S(x_T^j)$ & The recommendation for $x_T^j$ provided by source-trained $RS_S$ \\
    $\text{PL}^{\text{RS}}_{x_T^j}$ & The RS-based PL for $x_T^j$ \\
    $x_S^{(i)}$ & The category centroid for the $i$\textsuperscript{th} source category \\
    $RS_T(x_S^{(i)})^n$ & The top N recommendation for source centroid $x_S^{(i)}$ provided by target-trained $RS_T$ \\
    $\mathcal{L}_{ABR}$ & The adaptive bi-recommendation matching loss \\
    $P_T^j$ & The probabilistic output of $x_T^j$ from the classifier $C$ \\
    $P_T^{\mu}$ & The mean of all target probabilistic outputs \\
    ${p_T^{\mu}}^{(k)}$ & The $k$\textsuperscript{th} element of $p_T^{\mu}$ \\
    $\mathcal{L}_{DIV}$ & The diversity maximisation loss \\
    $\mathcal{L}_{TE}$ & The Tsallis Entropy loss \\
    $\alpha$ & The entropic index in $\mathcal{L}_{TE}$\\
    $\mathds{1}()$ & One hot vector \\
    $\text{PL}^{\text{NN}}$ & The neural network-predicted PL \\
    $\text{PL}^{\text{RS}}$ & The recommender system-decided PL \\
    $\text{PL}^{\text{SR}}$ & The source label relationship information-based PL \\
    $\text{PL}^{\text{TR}}$ & The target label relationship information-based PL \\
    $EK^{(K)}$ & The error knowledge of intrusion category $k$ \\
    $EK_0$ & The zero error knowledge vector \\
    $EK_{\mathcal{R}}^{(k)}$ & The reverse error knowledge of intrusion category $k$ \\
    $EK_{\mathcal{P}}^{(k)}$ & The previous error knowledge of intrusion category $k$ \\
    $\psi$ & The negative constant applied for $EK_{\mathcal{R}}^{(k)}$ \\
    $\phi$ & The negative constant applied for $EK_{\mathcal{P}}^{(k)}$ \\
    $\mathcal{L}_{EKL}$ & The error knowledge learning loss \\
    $\mathcal{L}_{SUP}$ & The source domain supervision loss \\
    $\mathcal{L}_{CE}$ & Cross entropy loss \\
    $\rho$, $\gamma$, $\delta$, $\tau$ & Balancing hyperparameter for $\mathcal{L}_{ABR}$, $\mathcal{L}_{EK}$, $\mathcal{L}_{DIV}$ and $\mathcal{L}_{TE}$, respectively \\ \Xhline{2\arrayrulewidth}
    \end{tabular}
    \vspace{-3mm}
\end{table}}

\end{document}